\documentclass[preprint,aps,nofootinbib,  11pt, superscriptaddress]{revtex4-1}
\usepackage{amsmath} \usepackage{graphicx} \usepackage{dcolumn}
\usepackage{bm} 
\usepackage{amssymb}
\usepackage{pstricks}

\usepackage{subfigure}
\usepackage{wasysym}

\begin{document}

\newlength{\figurewidth}
\setlength{\figurewidth}{0.6 \columnwidth}

\newcommand{\prtl}{\partial}
\newcommand{\la}{\left\langle}
\newcommand{\ra}{\right\rangle}
\newcommand{\dla}{\la \! \! \! \la}
\newcommand{\dra}{\ra \! \! \! \ra}
\newcommand{\we}{\widetilde}
\newcommand{\smfp}{{\mbox{\scriptsize mfp}}}
\newcommand{\smp}{{\mbox{\scriptsize mp}}}
\newcommand{\sph}{{\mbox{\scriptsize ph}}}
\newcommand{\sinhom}{{\mbox{\scriptsize inhom}}}
\newcommand{\sneigh}{{\mbox{\scriptsize neigh}}}
\newcommand{\srlxn}{{\mbox{\scriptsize rlxn}}}
\newcommand{\svibr}{{\mbox{\scriptsize vibr}}}
\newcommand{\smicro}{{\mbox{\scriptsize micro}}}
\newcommand{\scoll}{{\mbox{\scriptsize coll}}}
\newcommand{\sattr}{{\mbox{\scriptsize attr}}}
\newcommand{\sth}{{\mbox{\scriptsize th}}}
\newcommand{\sauto}{{\mbox{\scriptsize auto}}}
\newcommand{\seq}{{\mbox{\scriptsize eq}}}
\newcommand{\teq}{{\mbox{\tiny eq}}}
\newcommand{\sinn}{{\mbox{\scriptsize in}}}
\newcommand{\suni}{{\mbox{\scriptsize uni}}}
\newcommand{\tin}{{\mbox{\tiny in}}}
\newcommand{\scr}{{\mbox{\scriptsize cr}}}
\newcommand{\tstring}{{\mbox{\tiny string}}}
\newcommand{\sperc}{{\mbox{\scriptsize perc}}}
\newcommand{\tperc}{{\mbox{\tiny perc}}}
\newcommand{\sstring}{{\mbox{\scriptsize string}}}
\newcommand{\stheor}{{\mbox{\scriptsize theor}}}
\newcommand{\sGS}{{\mbox{\scriptsize GS}}}
\newcommand{\sBP}{{\mbox{\scriptsize BP}}}
\newcommand{\sNMT}{{\mbox{\scriptsize NMT}}}
\newcommand{\sbulk}{{\mbox{\scriptsize bulk}}}
\newcommand{\tbulk}{{\mbox{\tiny bulk}}}
\newcommand{\sXtal}{{\mbox{\scriptsize Xtal}}}
\newcommand{\sliq}{{\text{\tiny liq}}}

\newcommand{\smin}{\text{min}}
\newcommand{\smax}{\text{max}}

\newcommand{\saX}{\text{\tiny aX}}
\newcommand{\slaX}{\text{l,{\tiny aX}}}

\newcommand{\svap}{{\mbox{\scriptsize vap}}}
\newcommand{\sjam}{J}
\newcommand{\Tm}{T_m}
\newcommand{\sTS}{{\mbox{\scriptsize TS}}}
\newcommand{\sDW}{{\mbox{\tiny DW}}}
\newcommand{\cN}{{\cal N}}
\newcommand{\cB}{{\cal B}}
\newcommand{\br}{\bm r}
\newcommand{\cH}{{\cal H}}
\newcommand{\cHlt}{\cH_{\mbox{\scriptsize lat}}}
\newcommand{\sthermo}{{\mbox{\scriptsize thermo}}}

\newcommand{\bu}{\bm u}
\newcommand{\bk}{\bm k}
\newcommand{\bX}{\bm X}
\newcommand{\bY}{\bm Y}
\newcommand{\bA}{\bm A}
\newcommand{\bb}{\bm b}

\newcommand{\lintf}{l_\text{intf}}

\newcommand{\tXW}{{\text{\tiny XW}}}
\newcommand{\tRL}{{\text{\tiny RL}}}

\def\Xint#1{\mathchoice
   {\XXint\displaystyle\textstyle{#1}}%
   {\XXint\textstyle\scriptstyle{#1}}%
   {\XXint\scriptstyle\scriptscriptstyle{#1}}%
   {\XXint\scriptscriptstyle\scriptscriptstyle{#1}}%
   \!\int}
\def\XXint#1#2#3{{\setbox0=\hbox{$#1{#2#3}{\int}$}
     \vcenter{\hbox{$#2#3$}}\kern-.5\wd0}}
\def\ddashint{\Xint=}
\def\dashint{\Xint-}
\title{Microscopically based calculations of the free energy barrier
  and dynamic length scale in supercooled liquids: \\ The comparative
  role of configurational entropy and elasticity}

\author{Pyotr Rabochiy} \affiliation{Department of Chemistry,
  University of Houston, Houston, TX 77204-5003} 

\author{Peter G. Wolynes} \affiliation{Departments of Chemistry,
  Physics and Astronomy, and Center for Theoretical Biological
  Physics, Rice University, Houston, TX 77005} 

\author{Vassiliy Lubchenko} \affiliation{Department of Chemistry,
  University of Houston, Houston, TX 77204-5003}
\affiliation{Department of Physics, University of Houston, Houston, TX
  77204-5005}

\date{\today}

\begin{abstract}

  We compute the temperature-dependent barrier for
  $\alpha$-relaxations in several liquids, without adjustable
  parameters, using experimentally determined elastic, structural, and
  calorimetric data. We employ the random first order transition
  (RFOT) theory, in which relaxation occurs via activated
  reconfigurations between distinct, aperiodic minima of the free
  energy. Two different approximations for the mismatch penalty
  between the distinct aperiodic states are compared, one due to Xia
  and Wolynes (Proc. Natl. Acad. Sci. {\bf 97}, 2990), which scales
  universally with temperature as for hard spheres, and one due to
  Rabochiy and Lubchenko (J. Chem. Phys. {\bf 138}, 12A534), which
  employs measured elastic and structural data for individual
  substances. The agreement between the predictions and experiment is
  satisfactory, given the uncertainty in the measured experimental
  inputs. The explicitly computed barriers are used to calculate the
  glass transition temperature for each substance---a kinetic
  quantity---from the static input data alone. The temperature
  dependence of both the elastic and structural constants enters the
  temperature dependence of the barrier over an extended range to a
  degree that varies from substance to substance. The lowering of the
  configurational entropy, however, seems to be the dominant
  contributor to the barrier increase near the laboratory glass
  transition, consistent with previous experimental tests of the RFOT
  theory using the XW approximation. In addition, we compute the
  temperature dependence of the dynamical correlation length, also
  without using adjustable parameters. These agree well with
  experimental estimates obtained using the Berthier et al.(Science
  {\bf 310}, 1797) procedure. Finally, we find the temperature
  dependence of the complexity of a rearranging region is consistent
  with the picture based on the RFOT theory but is in conflict with
  the assumptions of the Adam-Gibbs and ``shoving'' scenarios for the
  viscous slowing down in supercooled liquids.

\end{abstract}

\maketitle


\section{Motivation}

There is little consensus, at present, on the detailed connection
between the molecular motions preceding the glass transition and
system-specific interactions, in actual glass-formers. The
uncertainties are such that wildly different hypotheses about the
underlying physics are still entertained\cite{biroli:12A301}, despite
much recent progress\cite{LW_ARPC}.  The ambiguity in complete
molecular understanding also prevents one from calculating the liquid
relaxation rates and the glass-forming ability for specific
substances, making materials design difficult. Of course, such
difficulties are part of the bigger challenge---still faced by
materials scientists of all stripes---that of the {\em \`{a} priori}
prediction of the structure and phase behavior of compounds with
arbitrary stoichiometries.

Despite these issues, a microscopic foundation to address
quantitatively the structure and dynamics of supercooled liquids has
been provided by the random first order transition (RFOT) theory, see
Ref.\cite{LW_ARPC} for a review. In this picture, the transport in the
liquid above its glass transition is realized by local, activated
reconfiguration events between long-lived, quasi-equilibrium aperiodic
structures.  It is the central notion of the ROFT theory that
sufficiently below the crossover to activated transport, these
reconfigurations take place by a process resembling {\em nucleation}
of a new aperiodic arrangement within the one presently existing,
according to the following free energy profile\cite{KTW, XW}:
\begin{equation} \label{Fr} F(r) = 4 \pi r^2 \sigma_0 (a/r)^{1/2} - (4
  \pi/3) (r/a)^3 T s_c,
\end{equation}
where the first term on the right hand side represents the mismatch
penalty between the initial and final configuration.  The scaling of
the penalty with the droplet size $r$ as $r^{3/2}$---not the familiar
$r^2$---implies, effectively, a curvature dependent surface tension
coefficient $\sigma = \sigma_0 (a/r)^{1/2}$\cite{KTW, Villain}. The
quantity $\sigma_0$ is the mismatch penalty at length scale $a$ equal
to the volumetric size of the chemically rigid molecular unit, or
``bead''\cite{LW_soft, BL_6Spin}. The second term on the right hand
side gives the bulk-driving force for the reconfiguration, which, in
equilibrium, is completely determined by the excess, configurational
entropy of the liquid, denoted here as $s_c$, per bead. The driving
force also has an enthalpic component during aging, where the initial
state has stored energy\cite{LW_aging}.

The free energy profile (\ref{Fr}) yields the following activation
barrier\cite{KTW, XW}:
\begin{equation} \label{FXW} \frac{F^\ddagger}{k_B T} =
  \frac{3\pi(\sigma_0 a^2/k_B T)^2}{s_c/k_B}.
\end{equation}
Furthermore, the size $r^*$ such that $F(r^*) = 0$ corresponds to a
region that typically has an alternative state.  Thus it represents
the cooperativity length for the reconfigurations. Often, one uses a
volumetric size $\xi \equiv (4\pi/3)^{1/3} r^*$. From Eq.~(\ref{Fr}),
it follows that
\begin{equation} \label{xi} \xi/a = (4\pi/3)^{1/3} \left( \frac{3
      \sigma_0 a^2}{T s_c} \right)^{2/3}.
\end{equation}

According to Eq.~(\ref{xi}), the cooperativity size is determined by a
competition between two factors: the free energy cost of extending the
interface between alternative aperiodic states and their multiplicity.
Eq.~(\ref{FXW}) shows that the temperature dependences of these two
competing factors determine the temperature dependence of the
activation barrier.

The configurational entropy can, in principle, be calculated for
actual substances starting from the force laws by using the classical
density functional theory\cite{HallWolynes_JPCB} or recent
replica-approaches\cite{mezard:1076, RevModPhys.82.789}. In practice,
making such estimates is quite challenging for molecules that are not
spherically shaped. At any rate, a good approximation to $s_c$ can be
determined by integrating the calorimetrically determined heat
capacity of the glass and the supercooled liquid\cite{MartinezAngell,
  Angell_NIST}.

The mismatch penalty, or the surface tension coefficient $\sigma_0$,
cannot be measured directly in the laboratory, but again a recipe
exists for calculating it from first principles. The DFT formalism
does suggest some simple approximations. Xia and Wolynes\cite{XW} (XW)
put forth a general argument, by which the mismatch penalty is
approximated by the free energy cost of ``localizing'' individual
particles during the crossover to the activated transport\cite{XW},
see also the discussion of Eq.~(9) in Ref.\cite{RL_sigma0}.  As such,
the penalty is largely determined by the kinetic pressure---which, in
turn, scales approximately linearly with the ambient temperature (it
does so exactly for hard spheres)---and the logarithm of the
vibrational displacement of a particle around its average position in
the aperiodic lattice. This yields the following, very simple formula
for the activation barrier $F^\ddagger$ in equilibrium\cite{XW,
  LW_soft, StevensonW}:
\begin{equation} \label{Fsc} \frac{F^\ddagger}{k_B T} \approx
  \frac{32.  k_B}{s_c}.
\end{equation}
In the assumption that the surface tension coefficient $\sigma_0$ in
Eqs.~(\ref{Fr}) and (\ref{FXW}) changes smoothly with temperature, the
above expression can be regarded as an asymptotic result that is
expected to work better the closer the system is to the ``ideal glass
transition'' at the Kauzmann\cite{Kauzmann} temperature $T_K$. Below
$T_K$, the glass is in a putative state for which the multiplicity of
the distinct aperiodic packings is subthermodynamic so that the
configurational entropy vanishes.  This ideal glass state is not
achievable by cooling directly, owing to the diverging relaxation
times; it is also predicted generally to be avoided by substances that
can crystallize into a periodic
structure\cite{SWultimateFate}. Although expected to be strictly
correct only asymptotically, the formalism leading to Eq.~(\ref{Fsc})
has given rise to dozens of quantitative predictions of seemingly
disparate glassy phenomena, without making use of adjustable
parameters. One of these is the ``fragility index'':
\begin{equation} \label{frag}
 m= \frac{\partial\log_{10}\tau}{\partial(T_g/T)}\Big|_{T=T_g}.
\end{equation}
By Eq.~(\ref{Fsc}), the XW estimate of the mismatch penalty
implies\cite{LW_soft, StevensonW}
\begin{equation} \label{mcp} m_\text{frag} \approx 20.7 \Delta
  c_p/k_B,
\end{equation}
where $\Delta c_p$ is the heat capacity jump at the glass
transition. The numerical constant 20.7 pertains specifically to
setting the laboratory glass transition to occur on the time scale
$10^5$~sec, see below for derivation. The constant would be (modestly)
different for other speeds of quenching.  The simple relation
(\ref{mcp}) between the kinetic and thermodynamic attributes of the
glass transition does indeed conform to observation, according to the
compilation of several dozen substances by Stevenson and
Wolynes\cite{StevensonW}, even though it appears to systematically
underestimate the fragility for the more fragile substances.

Dozens more predictions stemming from the XW estimate for the mismatch
energy factor $\sigma_0$ have been confirmed by experiment and
include, importantly, the size of the cooperative
reconfigurations\cite{GruebeleSurface, Spiess, RusselIsraeloff,
  CiceroneEdiger, Berthier}, see Refs.\cite{LW_ARPC, LW_RMP, L_JPCL,
  LW_Wiley} for reviews, and also more recent work in
Refs.\cite{Wisitsorasak02102012, RL_LJ, RL_Tcr, RL_sigma0,
  2013arXiv1305.0702W}. Note that the strict existence of the ideal
glass state is not necessary for the RFOT theory to be useful,
although quantitative predictions are expected to be the more accurate
the smaller the value of $s_c$, as already mentioned.  Recently,
ultrastable glasses have been produced by vapor deposition that are
characterized by values of $s_c$\cite{EdigerScience2007,
  EdigerPNAS2009, stevenson:234514}, that are significantly lower than
at the common laboratory $T_g$ achieved by cooling. The predictions of
the ROFT theory still seem to hold up since the limiting
configurational entropy predicted by surface motions in the RFOT
theory is confirmed.  It is also important to recognize that there is
a rigorous sense, in which the entropy crisis at $T_K$ exists as a
possible idealized fixed point, as has been shown in related spin
models\cite{MCT1} and via liquid theory approximations of Parisi and
coworkers\cite{mezard:1076, RevModPhys.82.789}.

Recently, Rabochiy and Lubchenko\cite{RL_sigma0} (RL) have estimated
the mismatch penalty in a different way, using a more detailed density
functional argument than XW employed. In their argument, the
activation barrier explicitly depends on the high-frequency elastic
moduli of the liquid and local coordination, via the structure factor
$S(k)$, in addition to the configurational entropy.  The calculations
of the mismatch penalty made by them for several substances were
limited to the vicinity of the glass transition because the elastic
constants and structure factor were not measured over sufficiently
extended temperature range.  Despite the complexity of the
approximation in Ref.\cite{RL_sigma0} and the potential ambiguities in
the experimentally measured inputs, the values of the mismatch penalty
obtained by RL near $T_g$ are quite close to the estimates found by
XW.

Here we use both the XW and RL derived expressions for $\sigma_0$ to
compute, for the first time, the entire temperature dependence of the
activation barrier for eight specific substances, over a finite
temperature range, without adjustable parameters. The calculation can
be justly called microscopic even though experimental input data are
used. The input parameters consist of experimentally determined
configurational entropy (XW and RL), high-frequency elastic constants,
and the structure factor (RL). These are all ``static'' quantities. In
addition, we will use the entropy of fusion to independently determine
the volumetric size of the chemically rigid molecular unit, or the
``bead.''  Unfortunately, the structure factor $S(k)$ is available
only at a single temperature for most of the substances in question.
We have thus been forced to use the $S(k)$ values measured at a
specific fixed temperature, usually near $T_g$, in the full
temperature range; it is possible to assess the resulting error in
some cases.  In addition to directly comparing to experimental
relaxation times, the predictions for the barriers are also tested by
computing the corresponding kinetic glass transition temperature from
the input static data and comparing these against experimental
data. Another quantity of interest we calculate is the fragility
coefficient. Finally, the same input data are used to estimate the
cooperativity length $\xi$ for the two approximations. These are then
compared to the values that are essentially experimentally determined
via a relation due to Berthier et al.\cite{Berthier}. We point out
that all of the input data here taken from experiment for actual
molecular systems can be obtained approximately using existing
theoretical formalisms for sufficiently simple systems like
Lennard-Jones spheres.

Of the four predicted quantities mentioned above, the absolute barrier
itself at a given temperature and the fragility, which is the rate of
barrier change with temperature, appear to be most sensitive to
uncertainty in the experimental inputs and potential errors of the
approximations. To separate the effects of these uncertainties on the
absolute value of barrier and the rate of its change with temperature,
we may rescale the barrier to match its known value at $T_g$, which is
equivalent to using one adjustable constant, say, the bead size. This
rescaling---which is in general, as we shall see, quite modest---has
the effect of eliminating error in the absolute value of the barrier
at the laboratory $T_g$: Using Eq.~(\ref{Fsc}) and
\begin{equation} \label{tau} \tau = \tau_0 e^{F^\ddagger/k_B T},
\end{equation}
one obtains $m_\text{frag} = (\log_{10} e) [F^\ddagger(T_g)/T_g]
\Delta c_p(T_g)/s_c(T_g) = \Delta c_p(T_g) (\log_{10} e) [\ln
\tau(T_g)/\tau_0]^2/32.$\cite{LW_soft, StevensonW}. The logarithm in
the latter expression is fixed by the timescale of the glass
transition, and so one is left only with assessing the relative
temperature {\em derivative} of the barrier.

While the complete temperature dependence of the rescaled barriers
does deviate somewhat from the experimental values, the discrepancy is
more noticeable for strong substances, for which the temperature range
between the crossover and glass transition is larger, as well as the
range to the ideal glass transition. In all cases, the RL predicted
barrier decreases with temperature consistently faster than does the
measured barrier. At the same time, the XW expression, which leaves
out the potential temperature dependence of the elastic constants,
owing to its hard sphere form, consistently yields a slower decay rate
of the barrier with temperature, as expected; Lubchenko and Wolynes
dealt with these barrier softening effects a decade ago\cite{LW_soft}.
Although the set of substances covered here is admittedly modest in
number, it does cover a wide chemical and fragility range suggesting
the systematic underestimation of the barrier found here is
meaningful. This, in turn, implies the RL expression is an adequate
stepping stone for further systematic improvements in the theory of
the absolute barriers.

The present work also allows us to determine the relative
contributions of the temperature dependence of the configurational
entropy and the temperature dependence of the elastic constants to the
barrier increase with lowering temperature, which has excited much
discussion\cite{BB_Wiley}. The elastic constant variation is central
to many non-RFOT based ideas about glassy dynamics. For instance,
recent work independently indicates there is an intimate connection
between the elastic constants and the translational symmetry breaking
in supercooled liquids\cite{PhysRevLett.103.025701, Yan16042013}. Some
works\cite{dyre:224108, klieber:12A544} go as far as to suggest that
the temperature dependence of the barrier is {\em exclusively} due to
changes of the shear modulus and nothing else.  In those scenarios,
the barrier and the shear modulus are taken to be simply proportional
to each other.  This simplicity requires there being a fixed
cooperativity length determined by some other physical principle for
each substance. In the purely elastic analysis, this size is then
regarded as a freely adjustable parameter.  The theory does not
provide a specific recipe for calculating it. In contrast, the
RFOT-based approach of RL dictates that the barrier depends on the
elastic constants in an intricate fashion, while the size factor is
predicted from the theory itself as in the XW implementation of
theory. Importantly, the RL approximation calls for the use of
high-frequency elastic moduli, i.e., on times shorter than the plateau
time scale of the stress relaxation function. Here we show that even
within the RL approximation, which appears to overestimate the
variation of the barrier with temperature, the entropic contribution
to the temperature dependence of the barrier progressively dominates
the elastic part with decreasing temperature.  Finally, we compute the
complexity of the rearranging region as a function of the
log-relaxation time as a means of testing the internal consistency of
the RFOT-based expressions for the relaxation barrier. We establish
that the latter expressions are indeed internally consistent, thus
lending further support to the present microscopic
picture. Conversely, we conclude that the experimentally determined
temperature dependence of the complexity is inconsistent with the
assumptions of several popular views on the structural glass
transition, such as the venerable Adam-Gibbs approach and several
realizations of the shoving model.

The article is organized as follows. In Section~\ref{Meth} we describe
the methodology by which the mismatch penalty is
computed. Section~\ref{Calc} contains the resulting calculations of
the barrier over a range of temperatures, the kinetic glass transition
temperature itself, the fragility coefficients, the cooperativity
lengths, and the complexities. In Section \ref{Disc}, we summarize and
discuss the present findings.

\section{Methodology}
\label{Meth}

Our goal is to compute, as functions of temperature, two basic
quantities predicted by the RFOT theory: the barrier for activated
reconfigurations, Eq.~(\ref{FXW}), and the corresponding cooperativity
length $\xi$, Eq.~(\ref{xi}). In this Section, we describe how we
determine the input quantities in Eqs.~(\ref{FXW}) and (\ref{xi}).

A simple estimate of the surface tension coefficient $\sigma_0$ at the
elemental length scale $a$ was obtained by Xia and Wolynes\cite{XW}:
\begin{equation} \label{sigmaXW} \sigma_0^\tXW = \frac{3}{4} (k_B
  T/a^2) \ln(\alpha a^2/\pi e) \approx 1.85 k_B T/a^2.
\end{equation}
where one assumes that the Lindemann ratio\cite{Lindemann,
  L_Lindemann} of the typical vibrational displacement to particle
spacing $\alpha^{-1/2}/a = 0.1$. This estimate is based on an assumed
sharp demarcation of localized and delocalized regions at the atomic
scale. The logarithm comes from the entropy cost of localization. The
temperature scaling would be appropriate for hard spheres, where all
free energies are strictly entropic in origin. Note substituting
Eq.~(\ref{sigmaXW}) into Eq.~(\ref{FXW}) yields Eq.~(\ref{Fsc}).  The
surface penalty, according to XW, can be thought of as, roughly, a
half of the missing ``bond'' energy for a particle allowing it to be
localized in an aperiodic crystal versus being part of a delocalized
liquid, in which many particle arrangements are allowed. An estimate
for this bond energy comes from the free energy cost of particle
localization when a uniform liquid freezes into an aperiodic
crystal\cite{XW}, see also discussion in Ref.\cite{RL_sigma0}. Near
the ideal glass transition itself, the latter free energy cost is
determined mostly by an entropic cost of localizing a particle from a
volume $a^3$---relevant for particle exchange---to the volume
prescribed by vibrations of magnitude $a/\sqrt{\alpha}$ around its
equilibrium position, hence Eq.~(\ref{sigmaXW}). Eq.~(\ref{sigmaXW})
is analogous to Turnbull's rule\cite{Turnbull} relating the surface
energy at a crystal-melt interface to the melting point.

More recently, Rabochiy and Lubchenko\cite{RL_sigma0} estimated
$\sigma_0$ also using a classical density functional argument but with
a different kind of reasoning. Their argument starts by assuming a
spatially broad transition between specific structural states. It
nevertheless concludes the interface is of molecular dimensions near
the laboratory $T_g$. RL's expression for the mismatch energy
explicitly includes details of the interactions and coordination via
the elastic constants and the structure factor $S(k)$:
\begin{align} \label{sigma1} &\sigma_0^\tRL = \frac{k_BT}{a^2} \left[
    \frac{Mc_l^2a^{-1}S'(\pi/a)}{12N_bk_BT} \right]^{1/2} \left\{2 -
    \left[ \frac{Mc_l^2}{8 \pi^2 k_B T N_b} \int_0^{\frac{\pi}{a}}
      S(k)k^2 dk \right]^{-1/2} \right\}, \text{ if } m \ge 2 k_B T/a^3
  \\ \label{sigma2} &\sigma_0^\tRL= \frac{Mc_l^2a^{-1}}{4\sqrt{6} \pi
    N_b} \left[S'(\pi/a) \int_0^{\frac{\pi}{a}} S(k)k^2 dk
  \right]^{1/2}, \text{ if } m < 2 k_B T/a^3
\end{align}
where the quantity $m$, given by the expression
\begin{equation} \label{m} m= \frac{Mc_l^2}{4\pi^2N_b}
  \int_0^{\frac{\pi}{a}}S(k)k^2dk,
\end{equation} 
reflects the free energy penalty for uniform variations of the order
parameter that describes the transition across the interface, see
immediately below. The quantity $M$ stands for the mass of the
stoichiometric unit, while the elastic constants enter via the
longitudinal speed of sound $c_l$. Finally, $S'(k) = dS/dk$.

The aforementioned order parameter, $\eta$, directly reflects the
relative amount of one specific aperiodic phase (1) and the other,
(2), whose individual, aperiodic spatially varying density profiles
are denoted by $\rho_1$ and $\rho_2$ respectively:
\begin{equation} \label{eta} \rho(\br)= \rho_1(\br)\frac{1
    -\eta(\br)}{2} + \rho_2(\br)\frac{1 +\eta(\br)}{2}.
\end{equation}
The values $\eta = \pm 1$ correspond to pure components 1 and 2
sufficiently far from the interface. The slow variation of $\eta$
yields the corresponding Landau-Ginzburg-Cahn-Hilliard
functional\cite{CahnHilliard} given by:
\begin{equation} \label{FCH} F\left[\eta(\br)\right] = \int
  \left[\frac{\kappa}{2} (\nabla\eta)^2 +f(\eta)\right]d^3\br.
\end{equation}
The expression for coefficient $\kappa$ at the square-gradient term
does not directly concern us in this paper (while it does enter the
expression for $\sigma_\tRL$). The bulk free energy cost for uniform
fluctuations of the order parameter are subject to a free energy cost
approximated as:
\begin{equation} \label{FetaEq}
f(\eta)=\left\{
\begin{array}{lc} 
  m(1+\eta)^2/2, & -1 \leq \eta \leq -\eta^\ddagger  \\ 
  k_B T/a^3, & -\eta^\ddagger < \eta < \eta^\ddagger, \text{ if } \eta^\ddagger > 0 \\
  m(1-\eta)^2/2, & \eta^\ddagger \leq \eta \leq 1  
\end{array} \right.
\end{equation}
where 
\begin{equation}
  \eta^\ddagger = \max[1 - (2k_B T/m a^3)^{1/2}, 0].
\end{equation}

The two expressions in Eqs.~(\ref{sigma1}) and (\ref{sigma2})
correspond to two somewhat distinct regimes for the interface between
alternative aperiodic structures. When the free energy cost for a
spatially uniform change of the order parameter exceeds a certain
threshold value (Eq.~(\ref{sigma1})), a multitude of (equally)
dissimilar liquid states should be accessed in the interface. This
would seem to correlate conceptually with further replica symmetry
breaking in the interface, a phenomenon seen in the replica field
theory approach of Dzero et al.\cite{PhysRevB.80.024204}. The
appearance of those other liquid states is reflected in the presence
of a flat portion in the bulk energy barrier in Eq.~(\ref{FetaEq})
just as in the replica instanton theory.  Otherwise,
(Eq.~(\ref{sigma2})), the transition occurs directly between the two
aperiodic states in question, so that the bulk energy barrier
$F(\eta)$ consists of two intersecting parabolas, i.e., the medium
responds mostly in an elastic fashion, except in a very narrow region.
We refer the reader to Ref.\cite{RL_sigma0} for further
details. Finally, note the $\pi$ in the denominator of
Eq.~(\ref{sigma2}) was inadvertently omitted during typesetting of the
original expression in Eq.~(53) of Ref.\cite{RL_sigma0}.

The final ingredient in implementing the RFOT theory using either the
RL calculation or XW calculation is the number of beads per
stoichiometric unit $N_b$, or ``bead count.'' This is needed to
evaluate the configurational entropy $s_c$ per bead from
Eq.~(\ref{FXW}) and the volumetric bead size $a$. Given the
experimentally determined entropy per stoichiometric unit $s_c^{(m)}$,
the entropy per bead is given by:
\begin{equation} \label{sc} s_c = \frac{s_c^{(m)}}{N_b}.
\end{equation}
The bead count $N_b$ can be estimated by dividing the entropy of
fusion of a substance by that for an interacting material of spherical
particles; the logic is that the degrees of freedom that freeze out
during crystallization are the same ones that would be pertinent for
the rearrangements in the liquid supercooled below its fusion
point\cite{LW_soft}. (This notion is reasonable except for plastic
crystals or other examples of system-specific local ordering, see
discussion of B$_2$O$_3$ below.)  Specifically, here we follow
Refs.\cite{LW_soft, StevensonW, RL_Tcr, RL_sigma0} and take Ar as such
a spherical particle material, whose fusion entropy is $1.68 k_B$ per
particle. This yields the expression:
\begin{equation} \label{Nb}
 N_b= \frac{\Delta H_m}{1.68k_BT_m},
\end{equation} 
where $\Delta H_m$ and $T_m$ are the enthalpy and temperature of
fusion of the substance respectively. Eq.~(\ref{Nb}) corresponds to
what we call the ``calorimetric'' bead count. For concreteness, we
assume that the number of beads per stoichiometric unit is
$T$-independent, except for case of B$_2$O$_3$, for chemical reasons
described below. Because the density generally depends on temperature,
the volumetric bead size $a = (v/N_b)^{1/3}$ is (weakly) temperature
dependent, where $v$ is the specific volume of the liquid.

\section{Calculation of the barrier and the cooperativity length}
\label{Calc}

\begin{figure}[h]
\centering
\subfigure{\includegraphics[width = 0.36\textwidth]{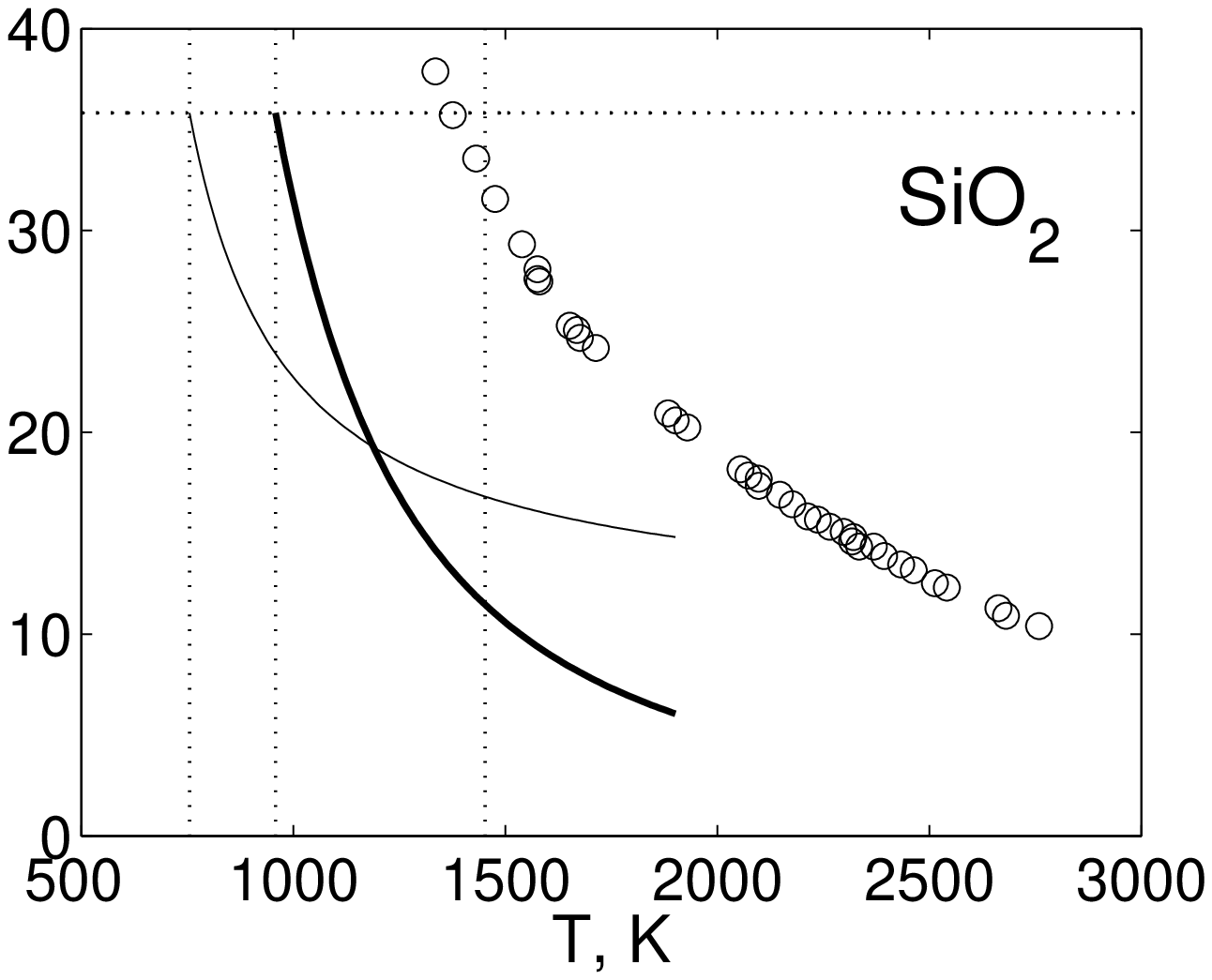}}\quad \: \: \: \: \:
\subfigure{\includegraphics[width = 0.36\textwidth]{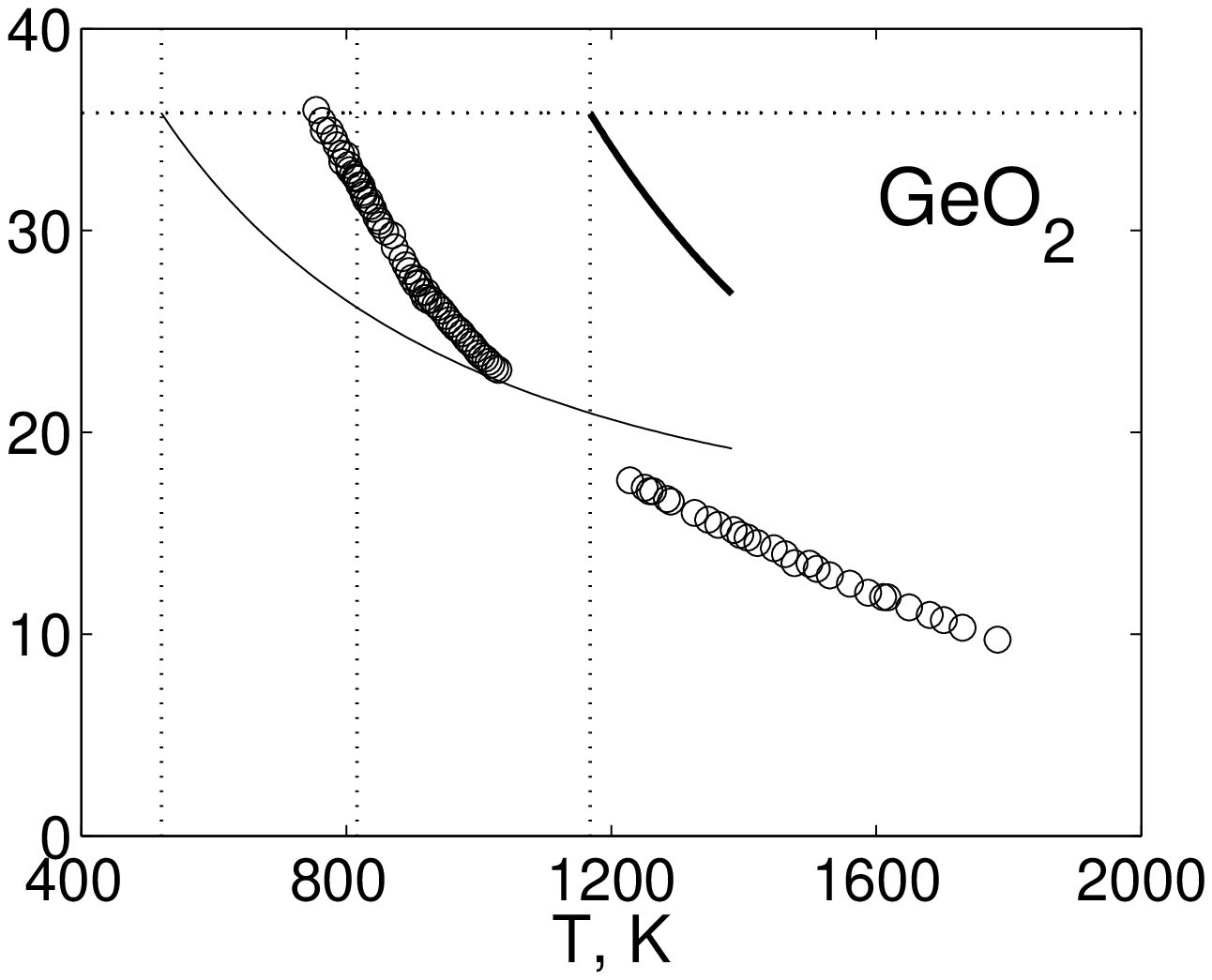}}\\
\subfigure{\includegraphics[width = 0.36\textwidth]{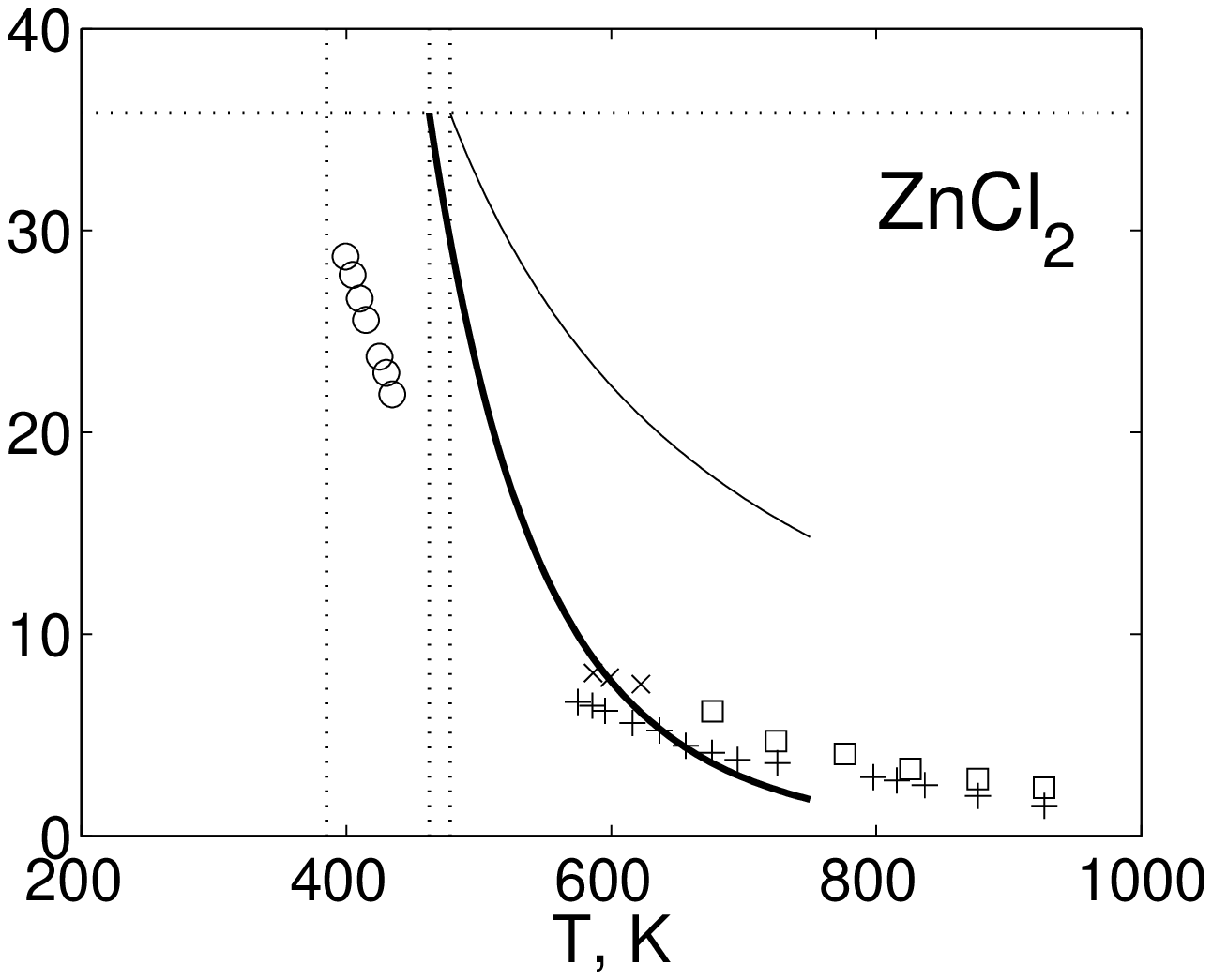}}\quad \: \: \: \: \:
\subfigure{\includegraphics[width = 0.36\textwidth]{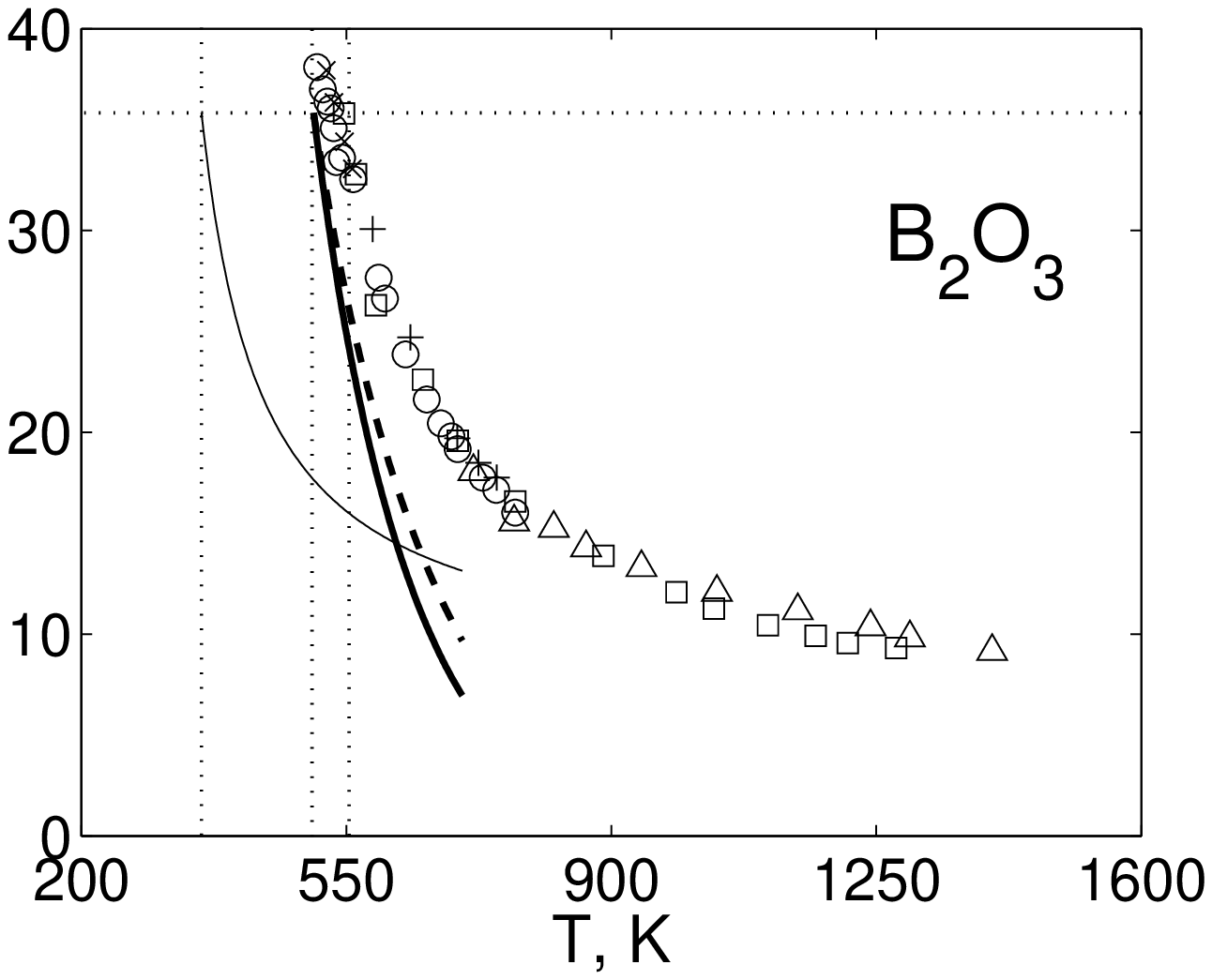}}\\
\subfigure{\includegraphics[width = 0.36\textwidth]{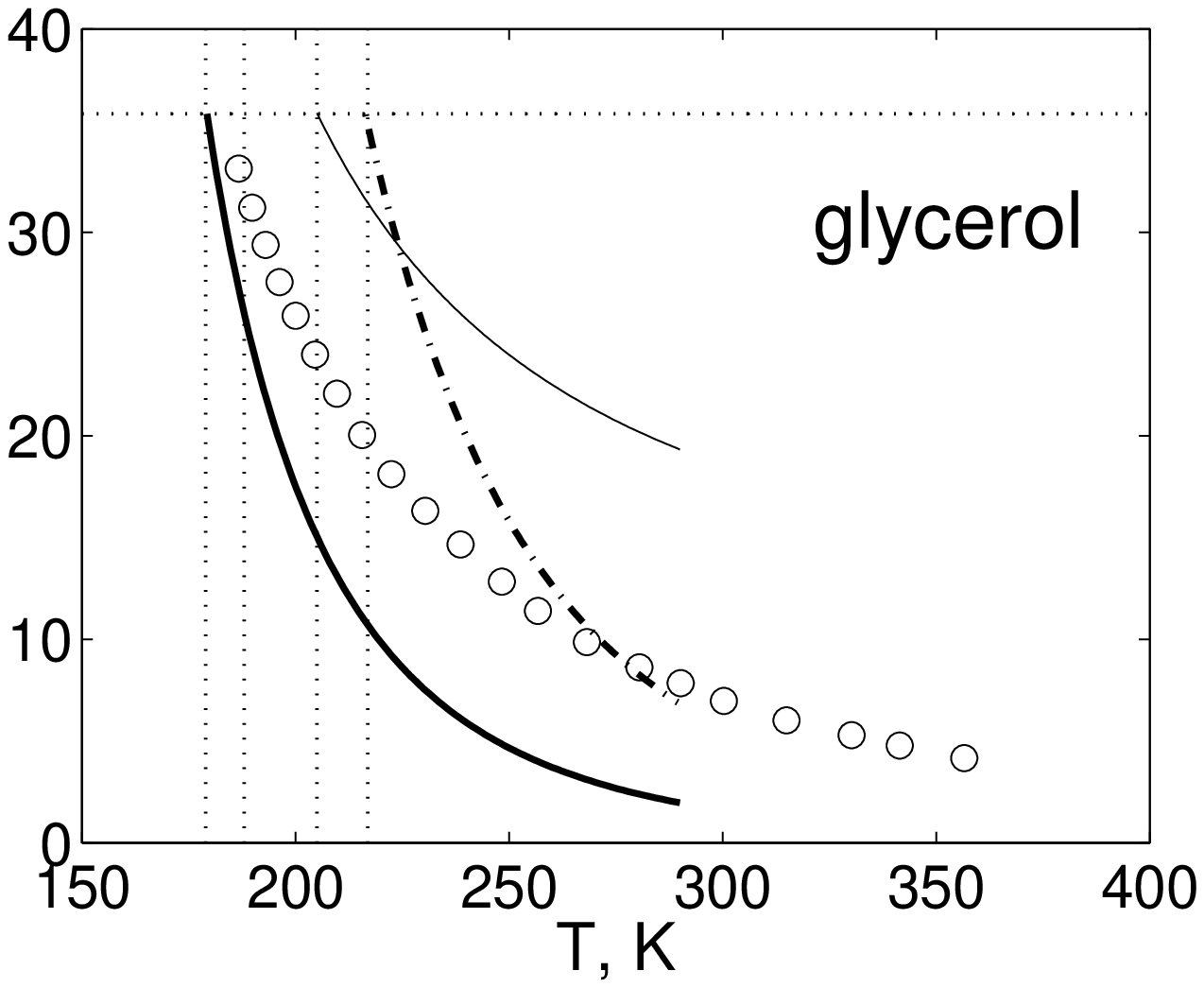}}\quad \: \: \: \: \:
\subfigure{\includegraphics[width = 0.36\textwidth]{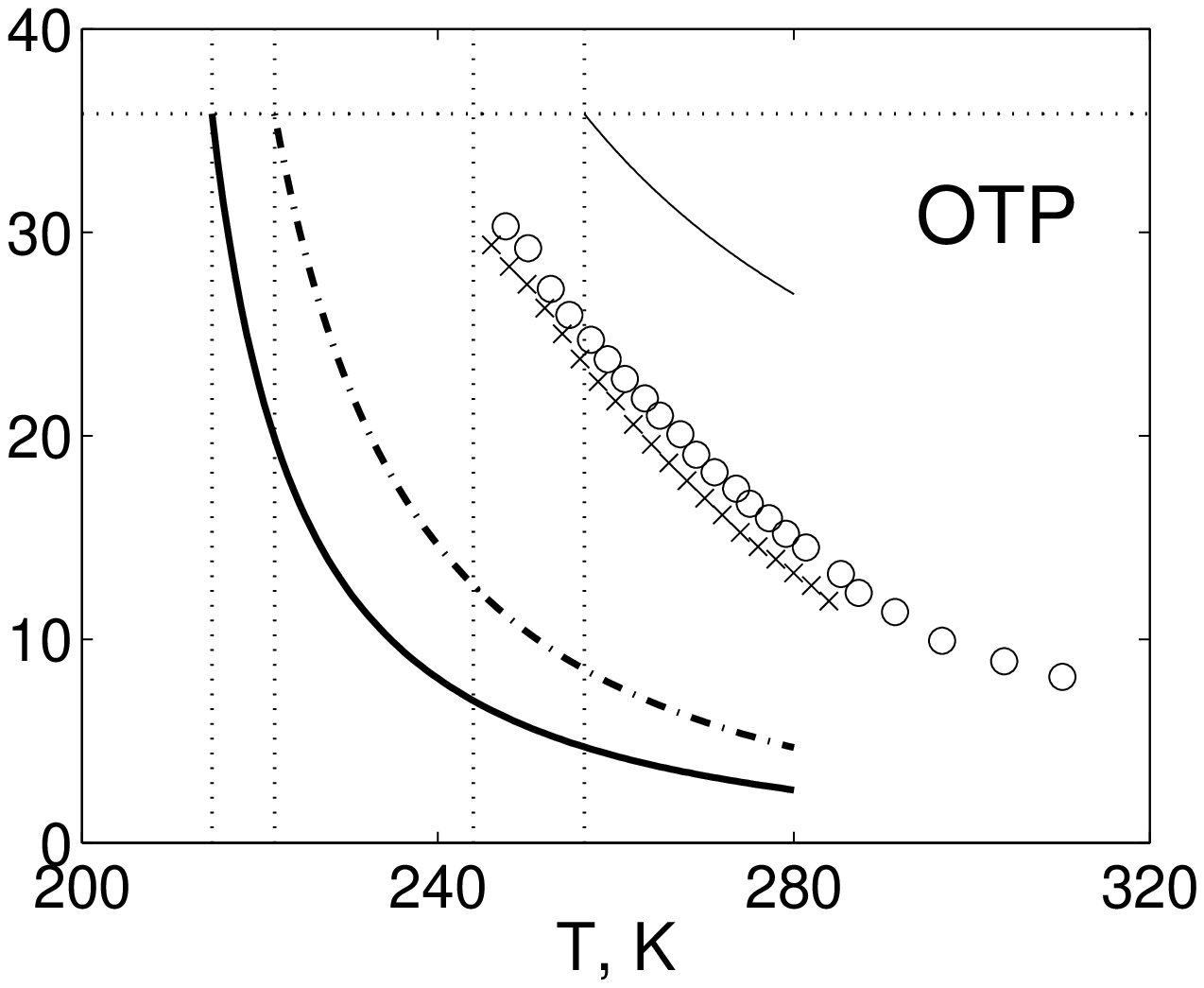}}\\
\subfigure{\includegraphics[width = 0.36\textwidth]{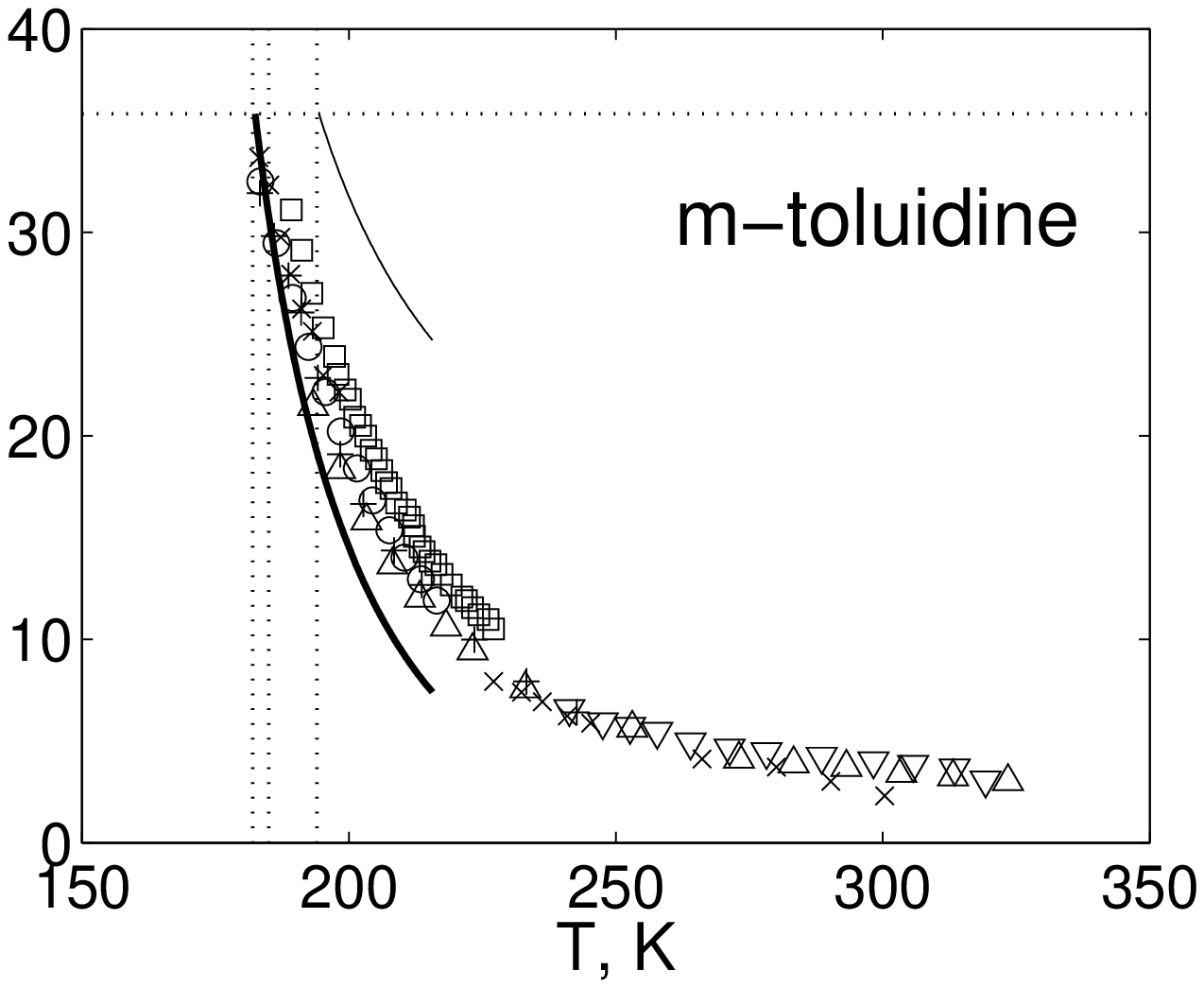}}\quad \: \: \: \: \:
\subfigure{\includegraphics[width = 0.36\textwidth]{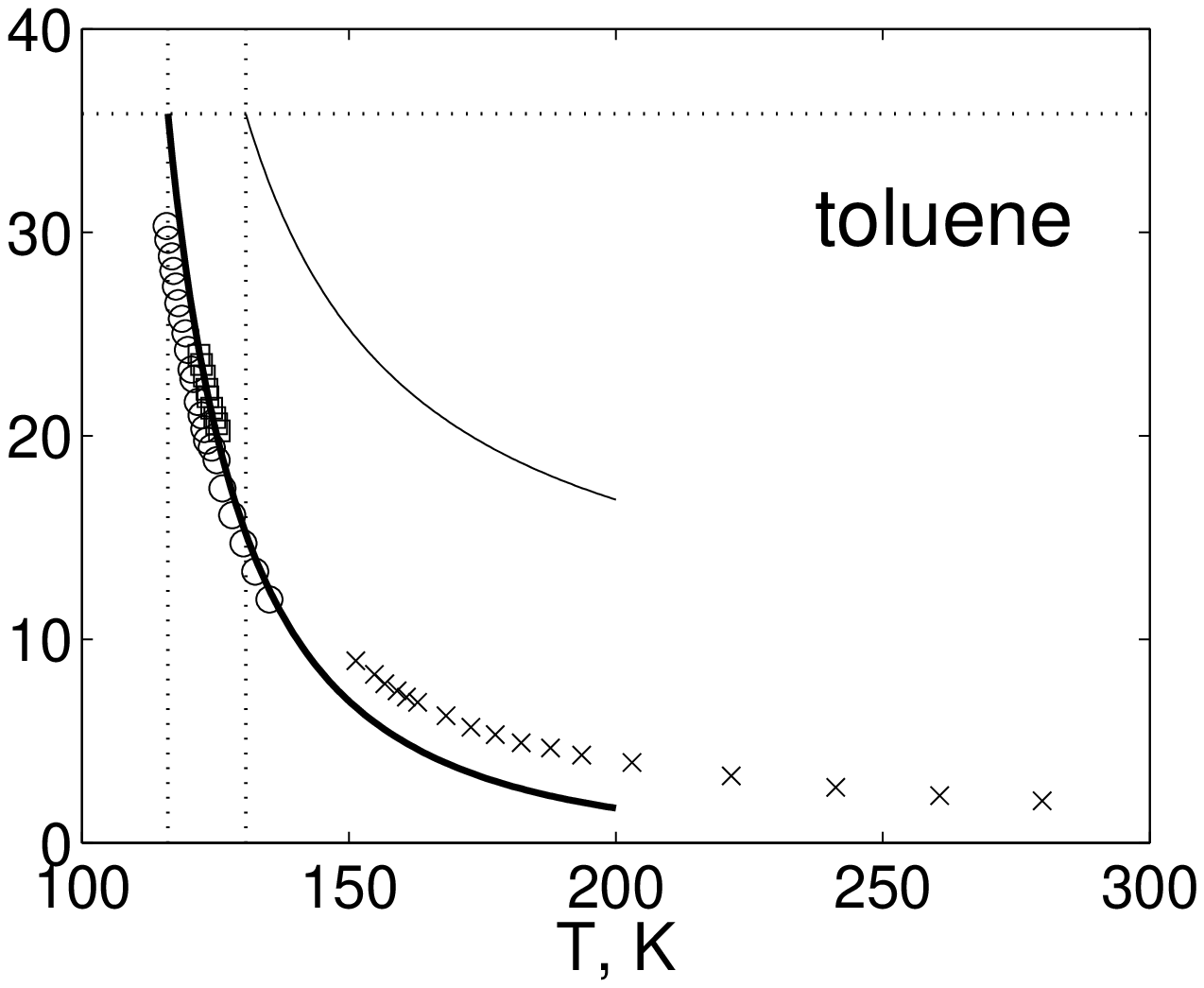}}\\
\caption{\label{Cal} The free energy barrier for $\alpha$-relaxation
  (divided by $k_B T$) as a function of temperature calculated using
  the calorimetric bead count. The XW approximation is shown with the
  thin solid line and RL approximation with thick lines, see the main
  text for detailed explanation. The symbols correspond to
  experimental data; different symbols denote dinstinct experiments.}
\end{figure}

Comparison of Eqs.~(\ref{sigmaXW}) with Eqs.~(\ref{sigma1}) and
(\ref{sigma2}) indicates that the RL approximation for the mismatch
penalty implies there is an additional source of temperature
dependence of the activation barrier, from Eq.~(\ref{FXW}), compared
with the purely entropic result from the XW hard sphere-based
approximation for $\sigma_0$.  We compare the predictions of the two
approximations for the barrier with experimental log-relaxation times
in Fig.~\ref{Cal}, for eight specific substances, for which all of the
relevant input data are available: SiO$_2$, GeO$_2$, ZnCl$_2$,
B$_2$O$_3$, glycerol, OTP, $m$-toluidine and toluene. The input data
include the temperature dependences of the configurational entropy,
elastic constants, and density.  Smooth fitting curves to
experimentally determined temperature dependences of the
configurational entropy $s_c^{(m)}$ (and the corresponding heat
capacity) per stoichiometric unit, for all eight substances under
consideration, have been kindly provided to us by Drs.~Zamponi and
Capaccioli. These fits were used by Capaccioli, Ruocco, and
Zamponi\cite{Capaccioli} to estimate the complexity $s_c (\xi/a)^3$ of
a rearranging region. Quantifying the complexity is of special
interest in the present context since it turns out to be a universal
function of the relaxation time in either RFOT-based result; we will
return to it later in the article.

We remind the reader that only the configurational entropy per bead is
needed to produce the predictions based on the XW approximation, while
the RL approximation needs structural and elastic input too. The
experimentally determined input parameters and/or references to the
original papers measuring them, in the case of temperature-dependent
quantities, are given in Table~\ref{Table1}. The substances are listed
in the Tables in the order of increasing of the fragility index $m$,
as experimentally determined. 
The bead counts from calorimetry $N_b$ and the corresponding bead
sizes $a$ are provided in Tables~\ref{TableRL} and \ref{TableXW}
together with other calculated quantities that are predicted by the
analysis.  Note in several instances, we needed to extrapolate the
elastic constants and calorimetric data to outside of the temperature
interval where these input data were available. The extrapolations for
the elastic data are shown in Fig.~\ref{cL} below, while those for the
configurational entropy are provided in the Supporting Information;
the latter also contains the fitting coefficients for both the elastic
constants and entropy.

In Fig.~\ref{Cal}, the symbols indicate the experimental values of the
free energy barrier (divided by $k_B T$), obtained from the measured
relaxation times. For the latter, we used data compiled by Capaccioli,
Ruocco, and Zamponi\cite{Capaccioli}, and other sources, see the
Supporting Information for details.  For this determination, the
high-$T$ limit of the relaxation time is taken to be $\tau_0 = 1$~ps:
$F^\ddagger/k_B T = \ln(\tau/\tau_0)$. In reality, although the
high-$T$ value of $\tau$ is usually in the vicinity of a picosecond,
as determined by the vibrational relaxation time, system-specific
deviations from the latter value may be present. Consequently, for
each order of magnitude worth of such deviation, the experimental
points on Fig.~\ref{Cal} would be off by $\ln 10 \approx 2.3$. The
thin solid line shows the free energy barrier computed with
Eq.~(\ref{FXW}) using the XW prediction (\ref{sigmaXW}) for
$\sigma_0$.

The thick lines in Fig.~\ref{Cal} show the barrier (divided by
temperature) computed using the RL prediction for $\sigma_0$,
Eqs.~(\ref{sigma1}) and (\ref{sigma2}). Some of the substances fall
into the regime corresponding to just one of these equations in the
full temperature range. In such cases, the corresponding equation
number is provided in Table~\ref{Cal}. Some substances obey
Eq.~(\ref{sigma1}) at low $T$ but make a change, at some temperature
above $T_g$, to the regime in which the uniform liquid states are
bypassed during reconfigurations so that Eq.~(\ref{sigma2}) now
applies. Under these circumstances, we provide, in Table~\ref{Cal},
the temperature of that change.

The graph for B$_2$O$_3$ contains a thick dashed line, which accounts
for the temperature dependence of its bead size. This matter has been
discussed extensively by two of us in Ref.\cite{RL_Tcr} and has to do
with a gradual ordering transition in B$_2$O$_3$, in which the rigid
molecular units at sufficiently high temperatures can be thought of as
BO$_3$ triangles, but at low temperatures they are essentially the
boroxol rings B$_3$O$_6$.

Last, but not least, all of the RL-based calculations mentioned so far
employ a temperature independent structure $S(k)$ measured near
$T_g$. Unfortunately, there is a lack of data at other temperatures
(see Table~\ref{Table1} for the specific temperature $T_S$ at which
$S(k)$ was measured for each substance).  For two of the liquids,
i.e., glycerol and OTP, we have also found $S(k)$'s measured at an
additional, higher temperature, see Table~\ref{Table1}.
We present the barrier values computed using the $S(k)$ measured at
these high temperatures as the dashed-dotted lines. We observe that
the computed barrier values, at high temperatures, move in the right
direction and, in the case of glycerol, almost coincide with the
measured value at the higher temperature at which $S(k)$ is known,
i.e., 295 K. This notion suggests that accounting for the temperature
dependence of the structural factor may be in fact important for
quantitative barrier estimates. Finally, we have not found $S(k)$ for
toluene measured at $T_g$, but only at a higher temperature {\em and}
higher pressure, which should be a reasonable substitute for the
$S(k)$ at $T_g$, as two of us argued elsewhere\cite{RL_sigma0}.

\begin{widetext}

\hspace{-15mm}

{

  \begin{table}[t]

\begin{tabular}{|c|c|c|c|c|c|c|c|c|c|c|c|c|} 
\hline &$M$, $10^{-25}$kg&$T_S$, K & $T_m$, K & $\Delta H$, kJ/mol &$c_l$ &$\rho$ &Eq. &Eq.(sc)  \\ \hline\hline
SiO$_2$\cite{Mei2008}        &1.00 &1452 &1995 &9.60 &\cite{0295-5075-57-3-375}  &\cite{PhysRevB.72.224201} &(\ref{sigma2}) &(\ref{sigma1})\\
GeO$_2$\cite{GeO2}           &1.74 &816  &1389 &17.1 &\cite{Youngman1997190}     &\cite{Dingwell1993}       &(\ref{sigma1}) &(\ref{sigma1})\\
ZnCl$_2$\cite{Zeidler2010}   &2.26 &385  &598  &10.3 &\cite{PhysRevE.63.041509}  &\cite{10.1063/1.1724847}  &(\ref{sigma1}) & $T_c=418$ K\\
B$_2$O$_3$\cite{Swenson1995} &1.16 &553  &723  &24.6 &\cite{PhysRevLett.62.2616} &\cite{macedo:3357}        &(\ref{sigma1}) &(\ref{sigma1})\\ 
glycerol\cite{Hanley1987}    &1.53 &183, 295  &291  &18.3 &\cite{4188596820090501,10.1063/1.1601608} &\cite{PhysRevB.70.054203} &(\ref{sigma2}) &(\ref{sigma2})\\
OTP\cite{Bartsch1993}        &3.82 &255, 314  &329  &17.2 &\cite{PhysRevE.68.011204,PhysRevE.63.061502} &\cite{PhysRevE.63.061502} &(\ref{sigma2}) &(\ref{sigma2})\\ 
m-toluid.\cite{Alba1998}     &1.78 &210  &242  &8.80 &\cite{dreyfus:7323} &\cite{aouadi:9860} & $T_c=210$ K & $T_c=212$ K\\ 
toluene\cite{Alba1998}       &1.53 &293* &178  &6.64 &\cite{rubio:4681}   &\cite{rubio:4681}  & $T_c=134$ K & $T_c=135$ K\\ 
  \hline    
  \end{tabular}

  \caption{ \label{Table1}
    The melting temperature $T_m$ is obtained from Ref.\cite{CRC} for all substances; 
    the fragility index $m$, glass transition temperature $T_g$, heat of fusion
    $\Delta H$ and heat capacity jump  $\Delta C_p$ at $T_g$ are from 
    Ref.\cite{10.1063/1.2244551} for all substances 
    except SiO$_2$, Refs.\cite{Bohmer,CRC,SiO2_GeO2_Tk}. The quantity  $T_S$
    is the temperature at which the structure factor $S(k)$ is known. The asterisk 
    at toluene's  $T_S$ indicates the structure factor is known at a higher pressure, 
    see text.  $c_l$ and $\rho$ 
    stand for the longitudinal sound speed and density respectively. The last two 
    columns indicate whether Eq.~(\ref{sigma1}) and (\ref{sigma2}) was used to 
    calculate the surface tension in the RL approach. If both equations have been
    used---Eq.~(\ref{sigma1}) at lower $T$ and Eq.~(\ref{sigma2}) at higher $T$---
    we indicate the temperature $T_c$ that separates the two regimes. }

\end{table}
}

\end{widetext}

\begin{widetext}
\begin{center}
\begin{table}[t]

  \begin{tabular}{|c|c|c|c|c|c|c|c|c|c|c|c|} 
  \hline   & $a$, \AA   & $a^\text{sc}$, \AA & $N_b$ & $N_b^\text{sc}$ & $m_{\text{exp}}$ & $m_{\text{th}}$  & $m_{\text{th}}^\text{sc}$ & $T_g^{\text{exp}}$, K & $T_g^{\text{th}}$, K & $\xi_{\text{exp}}$, nm & $\xi_{\text{th}}$, nm  \\ \hline \hline
   SiO$_2$ & 5.09 & 4.49                     & 0.34  & 0.50            & 20  & 47  & 35                                                  & 1452 & 958              & 2.91  & 2.42       \\
   GeO$_2$ & 3.79 & 4.37                     & 0.88  & 0.57            & 20  & 27  & 30                                                  & 816  & 1168             & 1.93  & 1.73       \\
  ZnCl$_2$ & 4.13 & 4.64                     & 1.23  & 0.87            & 30  & 82  & 94                                                  & 385  & 462              & 2.27  & 2.33       \\
B$_2$O$_3$ & 3.33 & 3.13                     & 1.58  & 1.81            & 36  & 68  & 76                                                  & 553  & 506              & 1.06  & 1.40       \\
 glycerol  & 2.95 & 2.89                     & 4.50  & 4.82            & 53  & 108 & 103                                                 & 188  & 179              & 0.82  & 1.85       \\       
     OTP   & 4.50 & 3.44                     & 3.74  & 8.41            & 81  & 355 & 134                                                 & 244  & 214              & 1.60  & 3.91       \\ 
m-toluid.  & 3.99 & 3.90                     & 2.60  & 2.79            & 98  & 174 & 160                                                 & 185  & 182              & 1.61  & 2.44       \\ 
 toluene   & 3.82 & 3.79                     & 2.67  & 2.73            & 103 & 137 & 139                                                 & 116  & 116              & 1.60  & 2.46       \\ \hline    
  \end{tabular}

  \caption{\label{TableRL} The quantities $a$, $N_b$, $m$, and $T_g$ 
    stand for the  bead size,  bead 
    count,  fragility index, and glass transition temperature  respectively. 
    The index ``th'' signifies the quantity was computed using 
    the Rabochiy-Lubchenko (RL)  approximation.
    The index ``sc'' signifies the corresponding quantities were
    evaluated using the ``self-consistent'' procedure for determining 
    the bead count explained in the text. 
  } 
\end{table}
\end{center}
\end{widetext}

\begin{widetext}
\begin{center}
\begin{table}[t]

\begin{tabular}{|c|c|c|c|c|c|c|c|c|c|c|c|} 
  \hline   & $a$, \AA & $a^\text{sc}$, \AA & $N_b$ & $N_b^\text{sc}$ & $m_{\text{exp}}$ & $m_{\text{th}}$ & $m_{\text{th}}^\text{sc}$ & $T_g^{\text{exp}}$, K & $T_g^{\text{th}}$, K & $\xi_{\text{exp}}$, nm & $\xi_{\text{th}}$, nm \\ \hline \hline
   SiO$_2$ & 5.09  & 3.98                  & 0.34  & 0.72            & 20  & 37 & 9                                                   & 1452 & 755             & 2.91  & 2.76        \\
   GeO$_2$ & 3.79  & 3.42                  & 0.88  & 1.20            & 20  & 11 & 11                                                  & 816  & 521             & 1.93  & 2.04        \\
  ZnCl$_2$ & 4.13  & 4.98                  & 1.23  & 0.70            & 30  & 26 & 47                                                  & 385  & 478             & 2.27  & 2.25        \\
B$_2$O$_3$ & 3.33  & 2.51                  & 1.58  & 3.53            & 36  & 55 & 16                                                  & 553  & 359             & 1.06  & 1.76        \\
 glycerol  & 2.95  & 3.20                  & 4.50  & 3.53            & 53  & 39 & 49                                                  & 188  & 205             & 0.82  & 1.60        \\  
     OTP   & 4.50  & 4.86                  & 3.74  & 2.97            & 81  & 61 & 79                                                  & 244  & 256             & 1.60  & 2.47        \\ 
m-toluid.  & 3.99  & 4.38                  & 2.60  & 1.97            & 98  & 71 & 95                                                  & 185  & 194             & 1.61  & 2.17        \\ 
 toluene   & 3.82  & 4.48                  & 2.67  & 1.65            & 103 & 50 & 98                                                  & 116  & 131             & 1.60  & 2.08        \\ \hline    
  \end{tabular}

  \caption{\label{TableXW} The quantities $a$, $N_b$, $m$, and $T_g$ 
    stand for the  bead size,  bead 
    count,  fragility index, and glass transition temperature  respectively. 
    The index ``th'' signifies the quantity was computed using 
    the Xia-Wolynes (XW)  approximation.
    The index ``sc'' signifies the corresponding quantities were
    evaluated using the ``self-consistent'' procedure for determining 
    the bead count explained in the text. 
  } 
\end{table}
\end{center}
\end{widetext}

We observe that for most of the substances the agreement between the
experiment and the theoretical predictions does leave some room for
improvement.  However, given that no adjustable parameters were used,
the agreement seems satisfactory, considering the input data are
purely static. To put things in perspective, we note that we are
unaware of any existing calculation that starts from the corresponding
information such as (high-frequency) elastic constants and structure
factor of a liquid and produces a figure for the crystal-nucleation
barrier with an accuracy comparable to that seen in Fig.~\ref{Cal}.
At any rate, we feel the present degree of agreement would justify
further studies that include more substances and suggests that the
present approximations provide a reasonable formal basis for further
systematic improvement.

Let us list possible errors that may arise in applying the present
formalisms to specific substances. Both RFOT based expressions for the
absolute barrier rely on accurate estimates of the bead size, by
Eqs.~(\ref{sigmaXW}-\ref{sigma2}), (\ref{sc}). As discussed
elsewhere\cite{ZLMicro2}, the calorimetric bead count is especially
likely to be off when local ordering is present, as in chalcogenide
alloys\cite{ZLMicro2} or the aforementioned boron
oxide\cite{RL_Tcr}. To get a sense of the issue, consider the fusion
entropy for chalcogenides. Almost independent of the stoichiometry,
the fusion entropy in these materials is about $1.5 k_B$ per {\em
  atom}---a typical value for ionic compounds\cite{CRC}---suggesting
there is roughly one bead per atom. Yet it is clear that in
As$_2$Se$_3$, for instance, the AsSe$_3$ pyramids are deformed little
during liquid motions, implying instead a bead per arsenic
atom\cite{ZLMicro2}. Now, the RL calculation is quite sensitive to the
precise value of the structure factor. As discussed in detail in
Ref.\cite{RL_sigma0}, the latter is subject to significant
experimental uncertainty for the relatively small values of $k$ in
question. Last but not least, the RL formula relies on accurate
measurement of the elastic constants, which becomes progressively more
difficult at increasing temperatures especially because of the problem
of separating the purely elastic from the relaxational part of the
stress response.

From the theoretical vantage point, we point out that neither the XW
nor RL formulas completely contains all known effects that have been
previously considered within the RFOT theory. Specifically, neglected
are the barrier softening effects that arise from approaching the mode
coupling temperature\cite{LW_soft}. The latter effects stem from the
spatial variation in the order parameter in the critical droplet near
the crossover; the resulting transition rate between distinct
aperiodic states is faster than what the simple nucleation-like
argument predicts. The XW expression ignores the softening
altogether. It would predict a finite nucleation rate even above the
mean field dynamical transition\cite{LW_soft}. The RL expression does
exhibit some degree of softening since it employs empirically
determined elastic constants.  The RL approach nevertheless disregards
several aspects of the instanton structure including the limited
applicability of the renormalized surface tension coefficient in
Eq.~(\ref{Fr}) close to the crossover\cite{LW_soft}.  Indeed, contrary
to the assumptions behind Eq.~(\ref{Fr}), the reconfigurations are
accompanied by string-like excitations close to $T_\scr$ implying,
among other things, that the interface is not thin\cite{SSW}.

A potential concern specific to the XW estimate for $\sigma_0$ is that
it should be most accurate near $T_K$, as already mentioned.  One of
the sources of ambiguity in the RL approach, on the other hand, is the
simplifying assumption on the vibrational spectrum, which was assumed
to be Debye-like with a sharp cutoff at $k = \pi/a$. In reality, the
cut-off should be soft; the $S'(k)$ derivative would be replaced by a
more complicated, integral expression under these circumstances, thus
requiring the knowledge of $S(k)$ at even lower values of
$k$. Finally, there is potential error in how one converts from the
atomic to the bead-wise structure factor.

\begin{figure}[h]
\centering
\includegraphics[width = 0.9\textwidth]{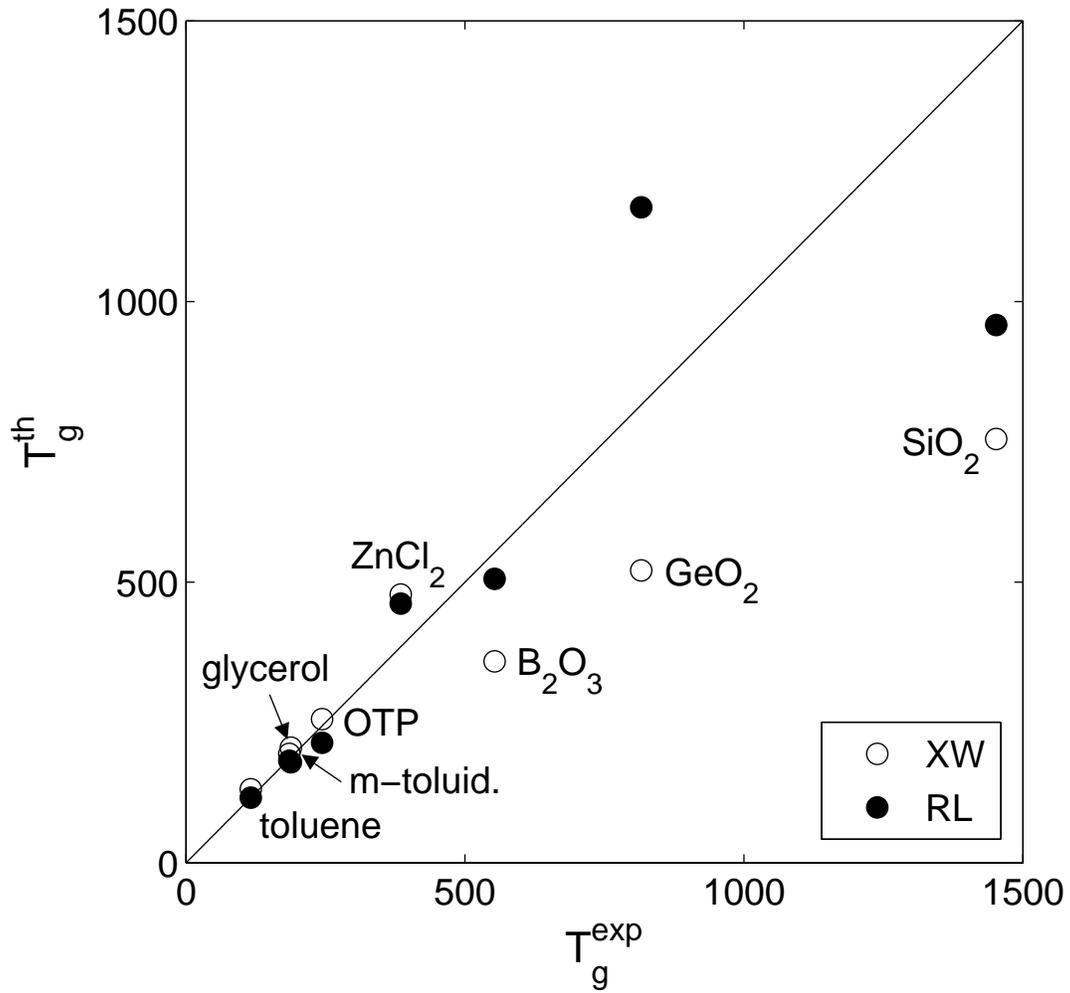}
\caption{\label{Tg} Theoretically predicted glass transition
  temperatures $T_g$, calculated using the XW and RL approximations
  for the mismatch penalty, plotted against experimental values. The
  theoretical temperatures are based on the barriers from
  Fig.~\ref{Cal} which rely on static input data, while the
  experimental values are kinetic quantities obtained from
  calorimetry.}
\end{figure}

It is instructive to provide a partial summary of the data in
Fig.~\ref{Cal} by plotting the glass transition temperatures, as
predicted by the theory, against their experimental values (they are
also tabulated in Tables~\ref{TableRL} and \ref{TableXW} for the RL
and XW-based estimates respectively). For the sake of argument, we
assume the time scale of the glass transition is $\tau_g = 1$hr, while
the vibrational relaxation time $\tau_0 = 10^{-12}$~sec. By
Eq.~(\ref{tau}), this implies $F^\ddagger(T_g)/T_g = \ln(3.6 \cdot
10^{15}) \approx 35.8$.  Fig.~\ref{Tg} shows the result of the
calculation. We observe that the predicted glass transition
temperatures agree well with the experimental values, despite the
sometimes large discrepancy between theory and experiment for the
rates themselves in Fig.~\ref{Cal}. Note that in Fig.~\ref{Tg} and the
Tables we use the experimental $T_g$ as determined calorimetrically,
instead of using relaxation data, and so one should expect some
discrepancy between the relaxation time-based apparent $T_g$, as
inferred from Fig.~\ref{Cal}, and the calorimetric $T_g$, which is
indicated by a vertical line on the same figure. This discrepancy is
generally present because of variations in the quench rate, sample
purity, detailed fitting of the DSC curve, etc.

\begin{figure}[h]
\centering
\includegraphics[width = 0.9\textwidth]{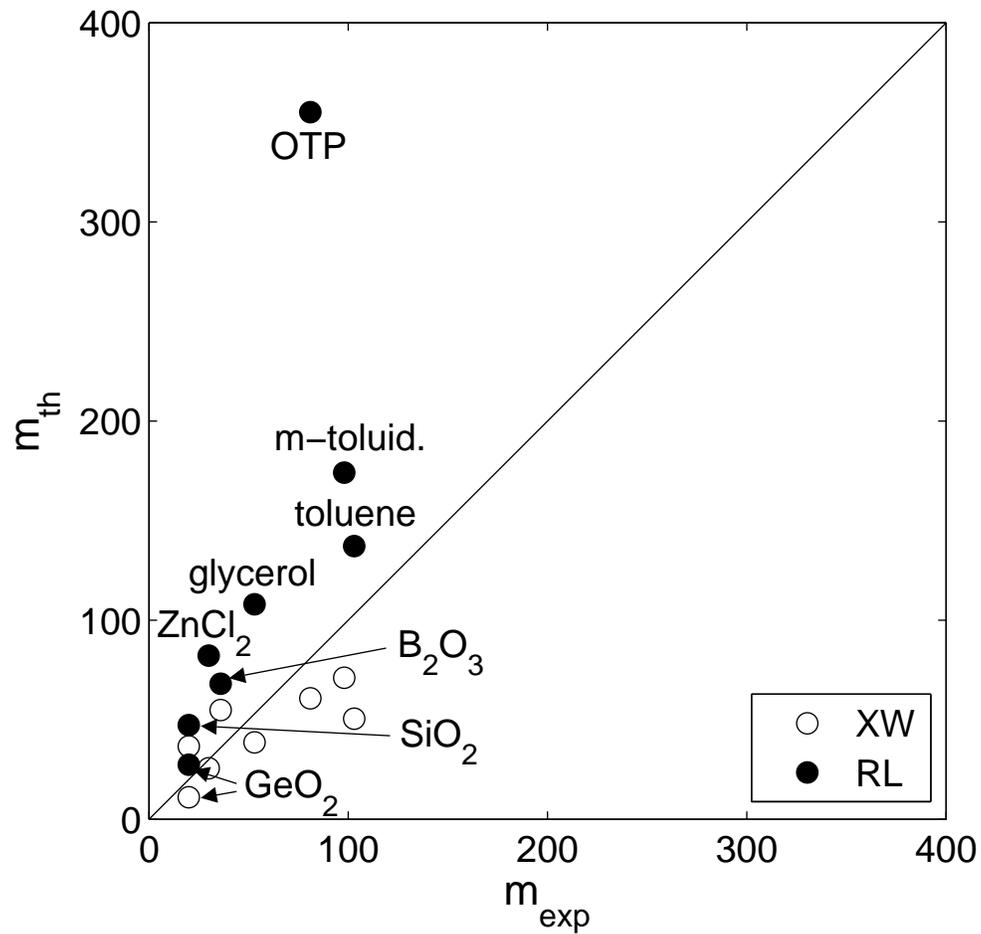}
\caption{\label{mCal} The fragility indices for several substances,
  Eq.~(\ref{frag}), corresponding to the barriers from Fig.~\ref{Cal}
  plotted against their experimental values.}
\end{figure}

Incidentally, one can compute the slopes of the curves from
Fig.~\ref{Cal} at the glass transition temperature, in the form of
fragility indices, see Eq.~(\ref{frag}). We emphasize that the
theoretically predicted fragility indices shown in the Tables were
computed at the predicted $T_g$, not at the experimental glass
transition temperature.  Fig.~\ref{mCal} shows this measure of liquid
fragility plotted against its experimental value, for the eight
substances in question. We note that both the RL and XW approximation
show a good correlation with measurement, except for the RL prediction
for OTP, to be commented on shortly.

\begin{figure}[h]
\centering
\subfigure{\includegraphics[width = 0.36\textwidth]{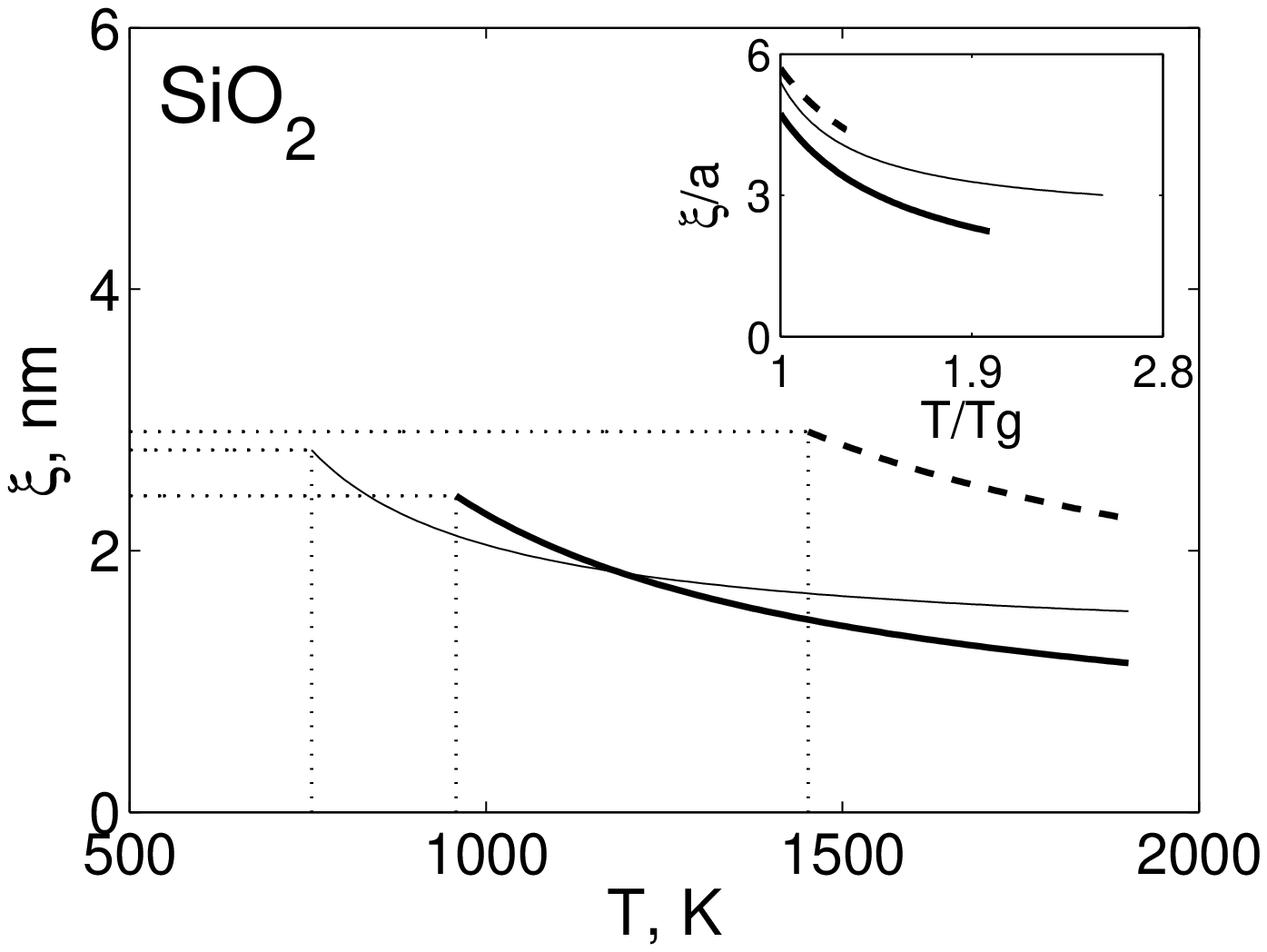}}\quad \: \: \: \: \:
\subfigure{\includegraphics[width = 0.36\textwidth]{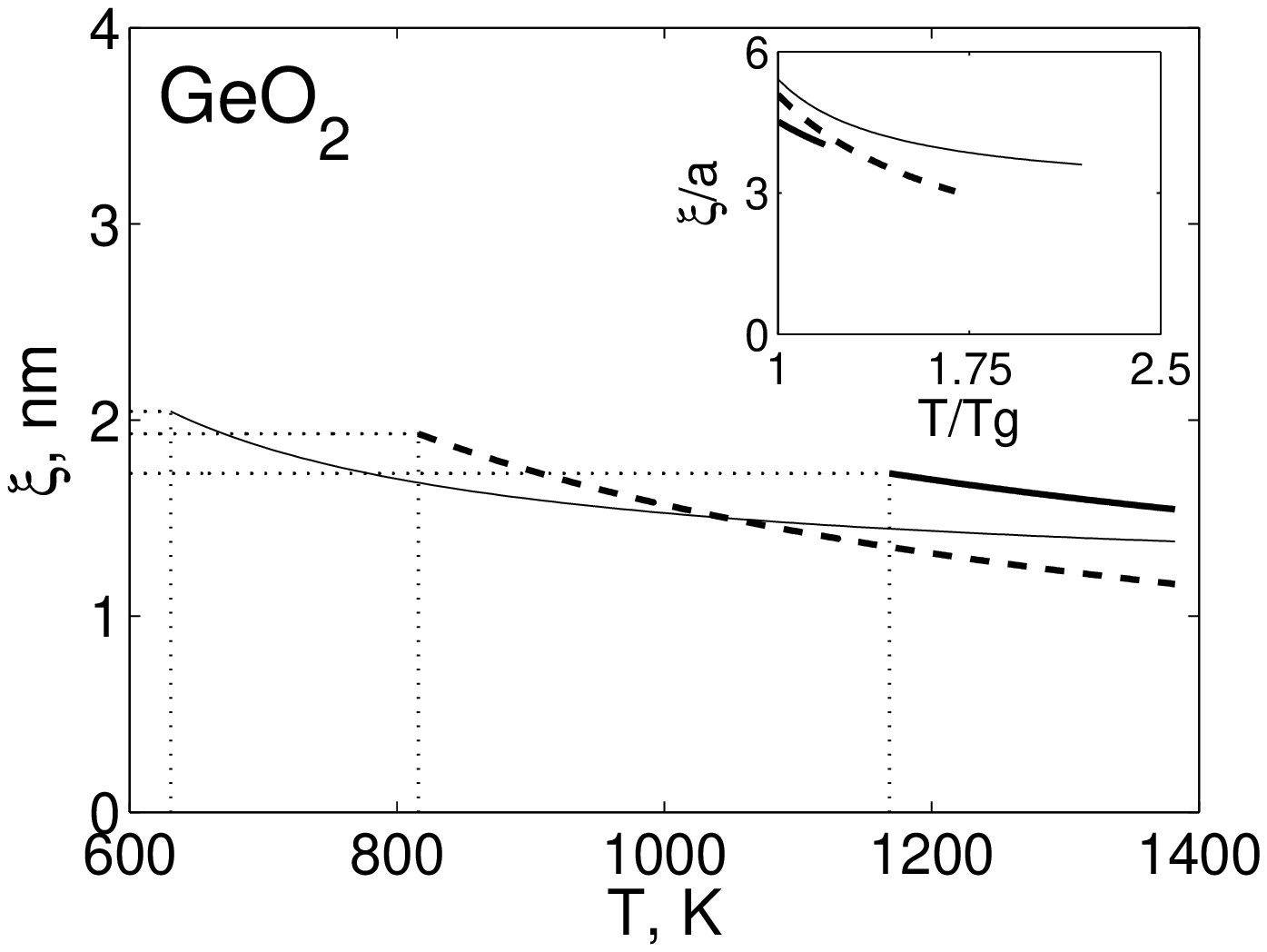}}\\
\subfigure{\includegraphics[width = 0.36\textwidth]{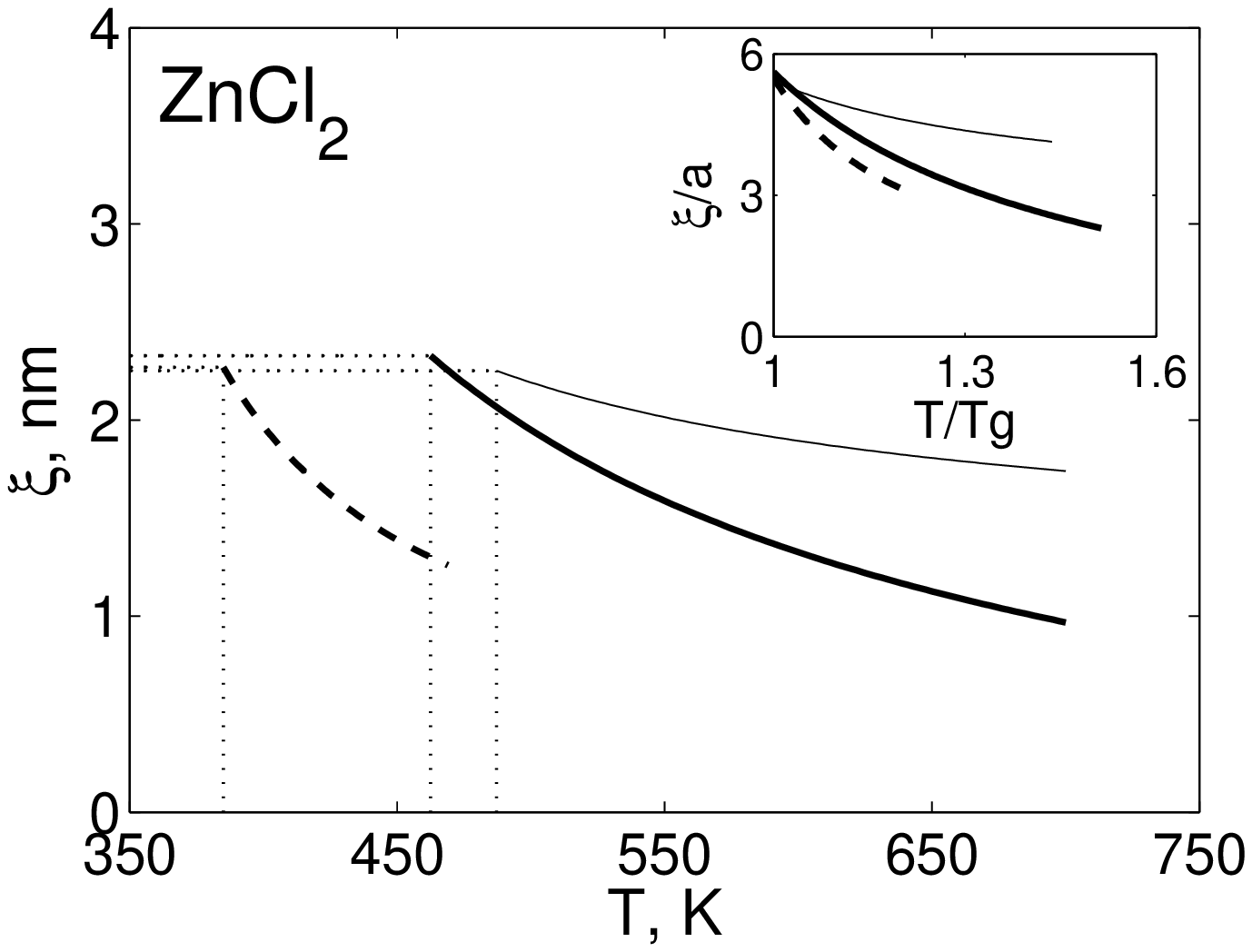}}\quad \: \: \: \: \:
\subfigure{\includegraphics[width = 0.36\textwidth]{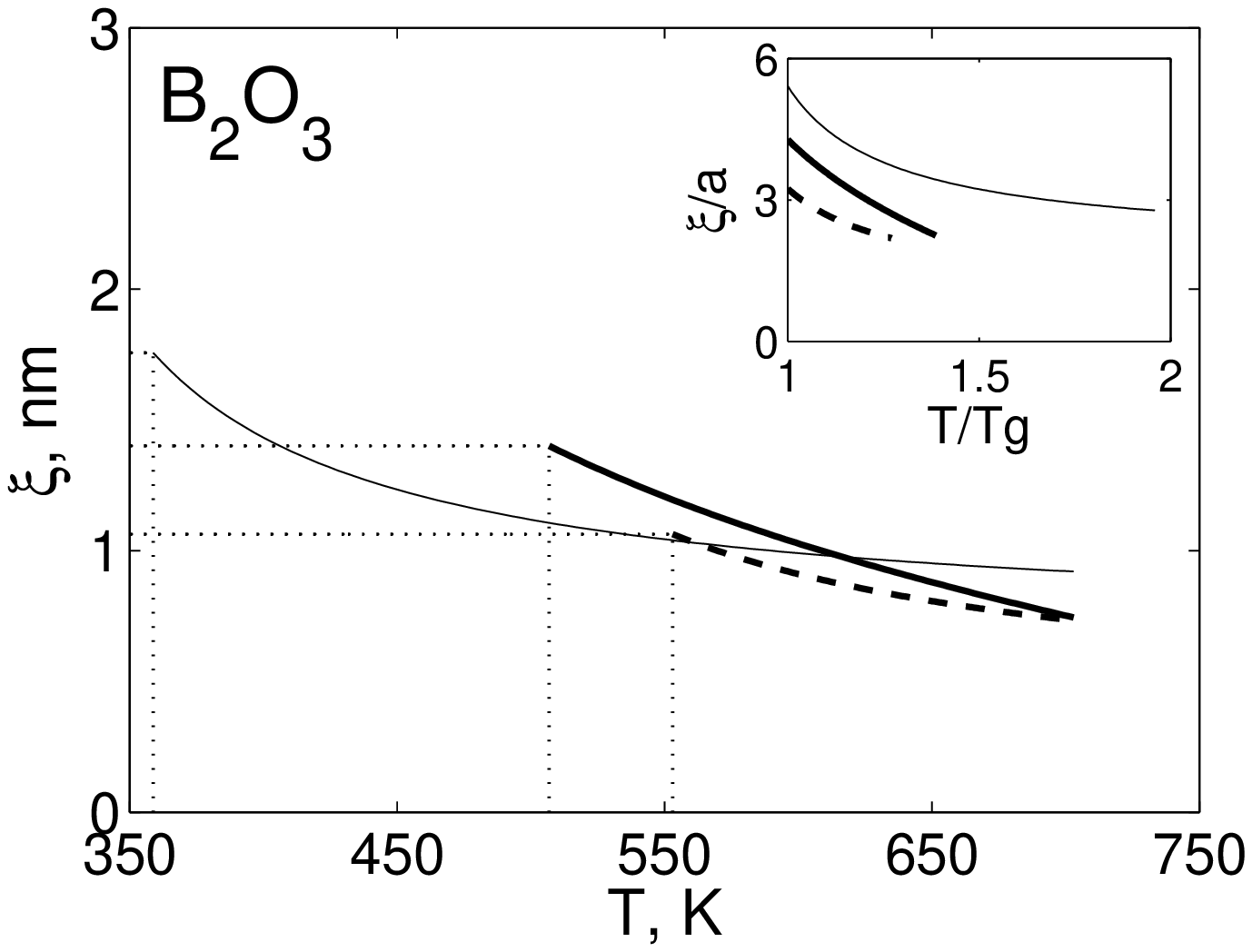}}\\
\subfigure{\includegraphics[width = 0.36\textwidth]{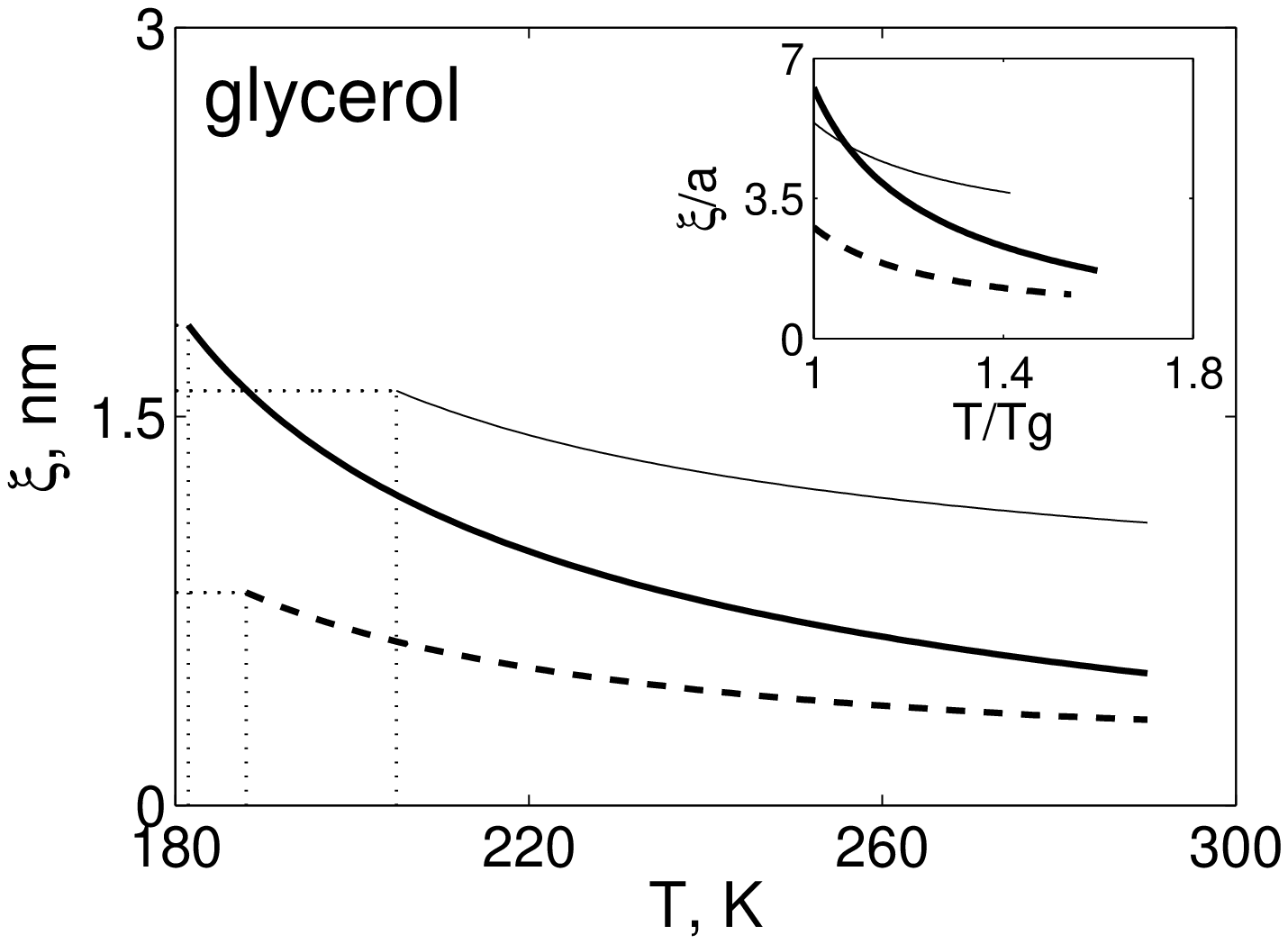}}\quad \: \: \: \: \:
\subfigure{\includegraphics[width = 0.36\textwidth]{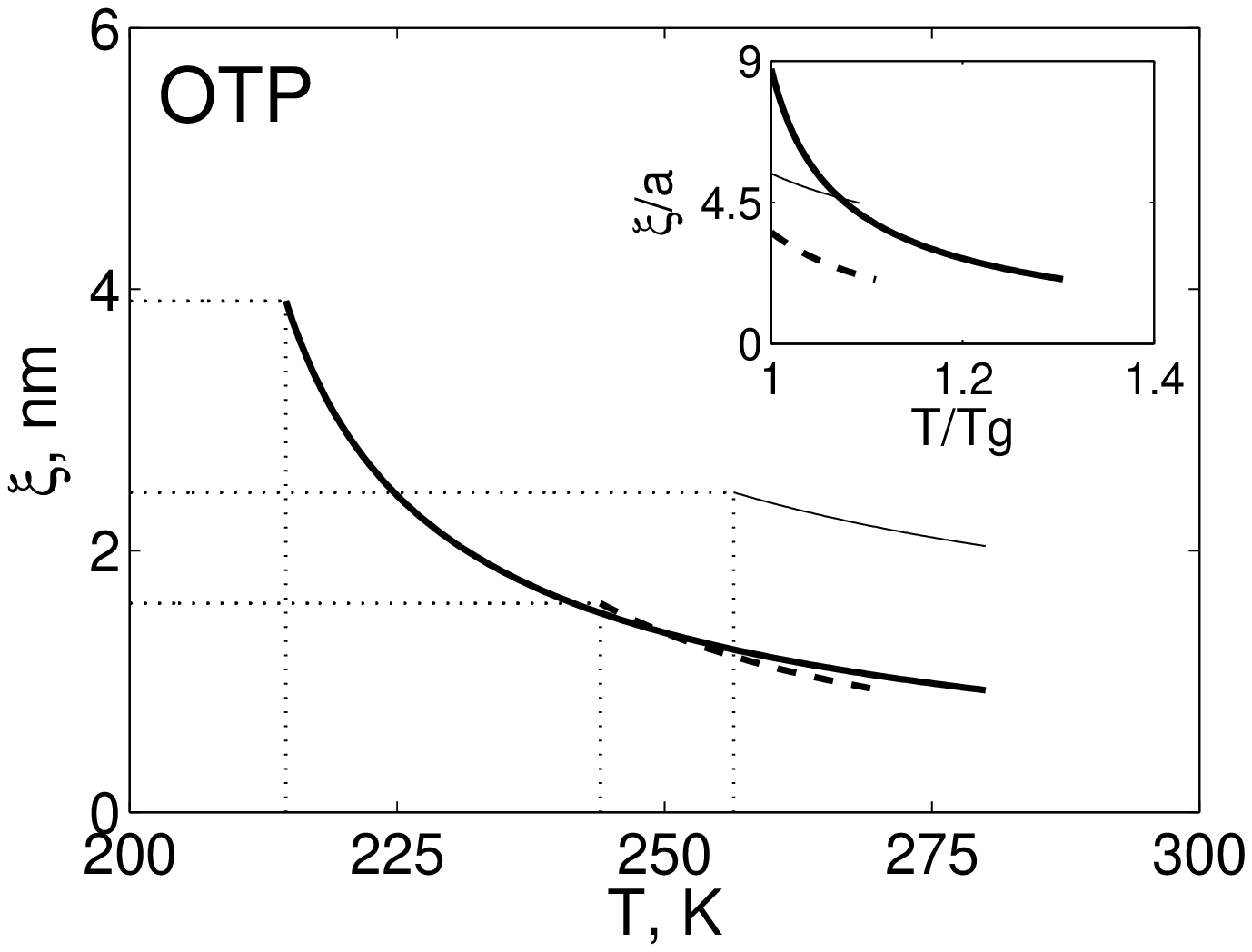}}\\
\subfigure{\includegraphics[width = 0.36\textwidth]{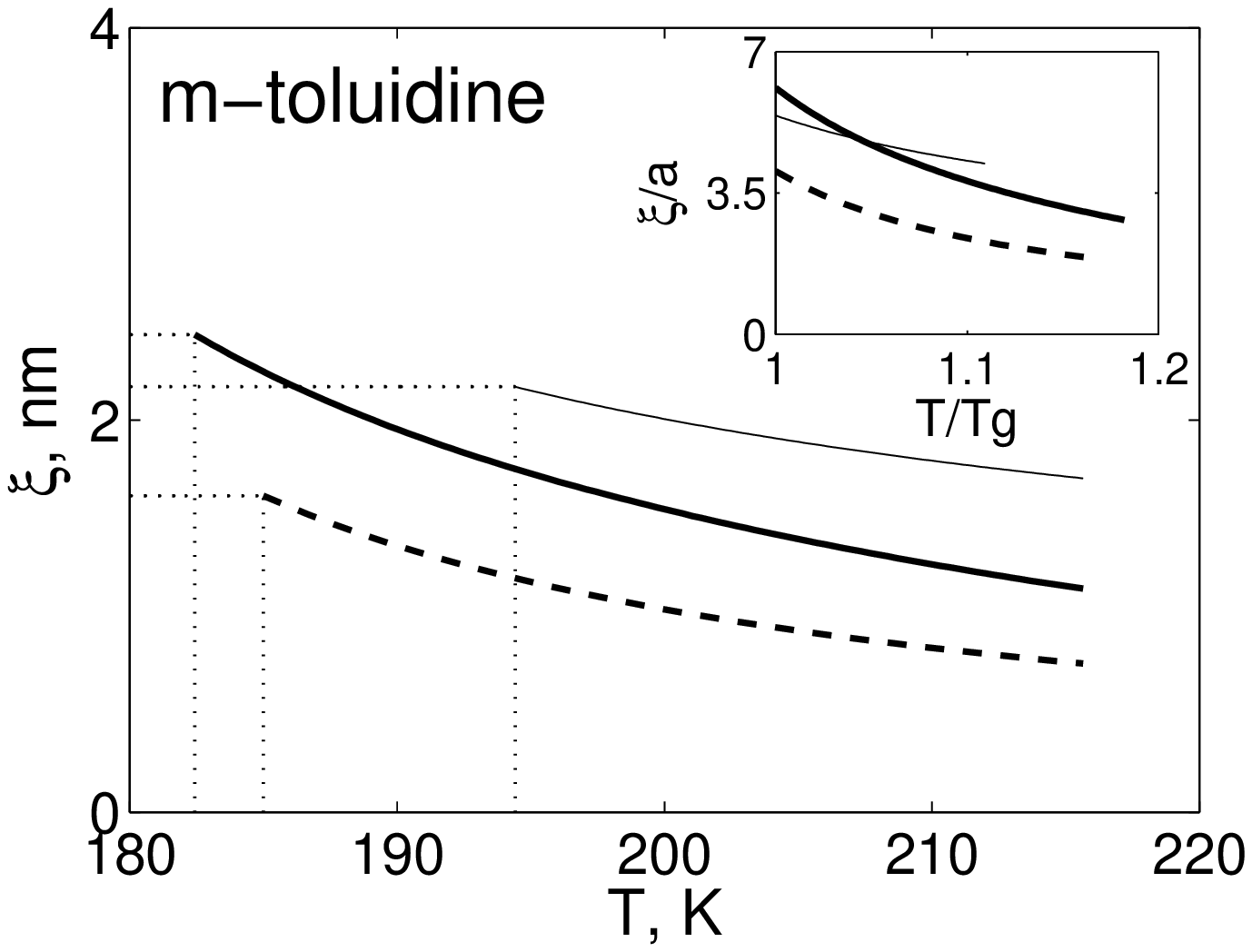}}\quad \: \: \: \: \:
\subfigure{\includegraphics[width = 0.36\textwidth]{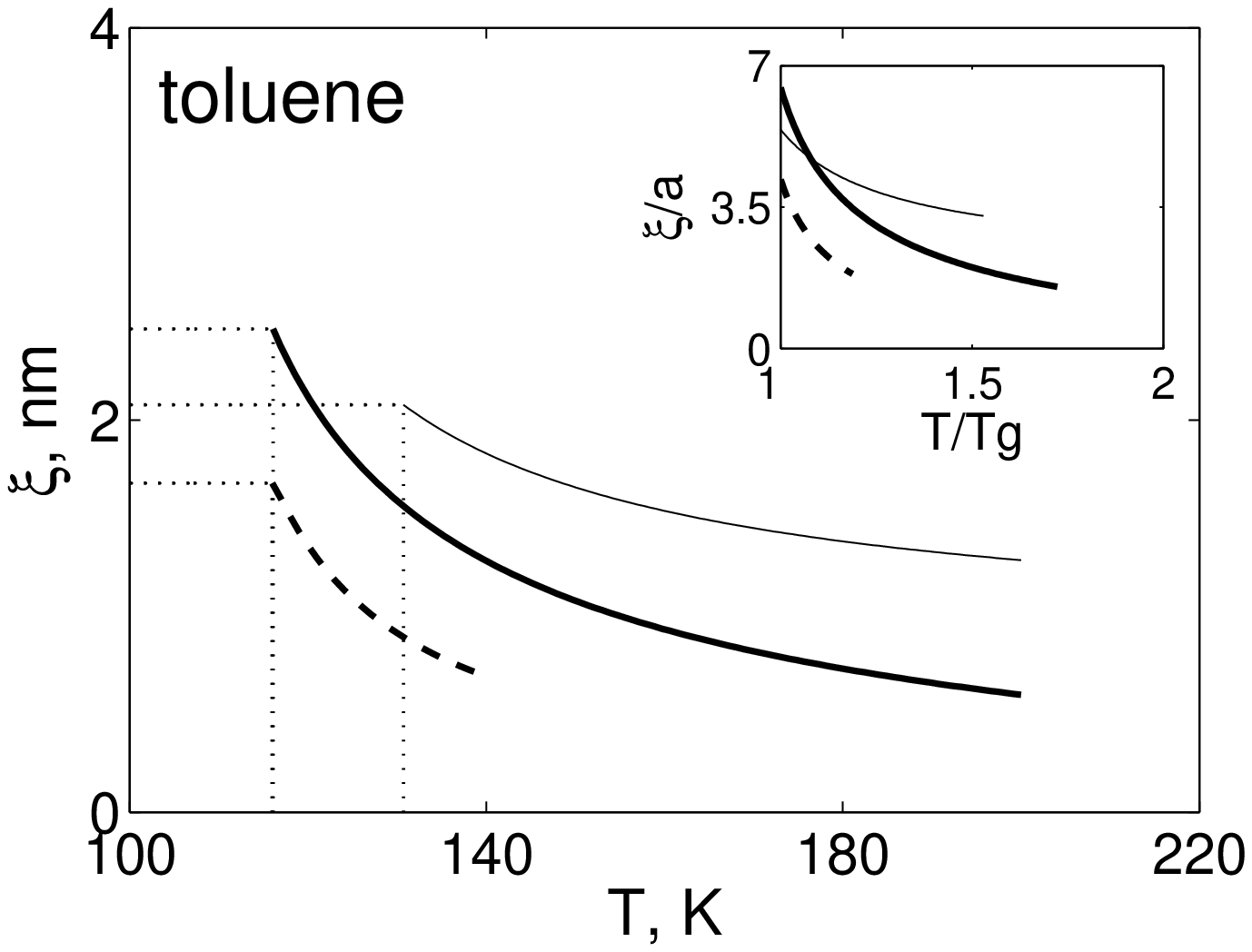}}\\
\caption{\label{xiCal} Comparison of the RFOT-based predictions for
  the cooperativity length $\xi$ as a function of temperature from
  Eq.~(\ref{xi}) (solid lines) with the ``experimental'' cooperativity
  length (dashed line) determined according to the procedure by
  Berthier et al.\cite{Berthier}, see Eq.~(\ref{xiBerthier}).  The
  thick and thin solid lines correspond to the RL and XW approximation
  for the mismatch penalty respectively. The vertical dotted lines
  indicate the respective glass transition temperatures for each set
  of data, as in Fig.~\ref{Cal}. The horisontal dotted lines are added
  as a visual aid. The insets are lengths scaled by the bead size and
  refer to temperature scaled by the respective glass transition
  temperatures.}
\end{figure}

Next, in Fig.~\ref{xiCal}, we show the temperature dependence of the
cooperativity length $\xi$ from Eq.~(\ref{xi}). The thin and thick
solid lines show the result of the XW and RL based calculations
respectively. The respective temperature intervals for the RL and XW
lengths correspond to the temperature intervals in Fig.~\ref{Cal};
they do not coincide because the predicted glass transition
temperatures are different for the two approximations.  Now, the
dashed line shows the dynamical correlation length $\xi$ computed
according to the procedure of Berthier et al.\cite{Berthier}:
\begin{equation} \label{xiBerthier} \xi/a = \left\{ \frac{1}{\pi}
    \left[ \frac{\beta}{e} \frac{\prtl \ln \tau}{\prtl \ln T}
    \right]^2 \frac{k_B}{\Delta c_p} \right\}^{1/3},
\end{equation}
where $\beta$ is the exponent in the stretched exponential relaxation
profile: $e^{-(t/\tau)^\beta}$. As already mentioned, we have used the
calorimetric data of Capaccioli, Ruocco, and
Zamponi\cite{Capaccioli}. In addition we have used their fitting forms
for the relaxation times and, when available, the stretching exponent
$\beta$. The corresponding fits are shown in the Supporting
Information.  For three of the substances---glycerol, ZnCl$_2$, and
B$_2$O$_3$---the temperature dependence of $\beta$ was available,
while for the rest of the substances, a constant value for this
quantity, as measured near $T_g$, was adopted for the lack of data for
$\beta$ in the extended temperature range just as was done by
Capaccioli et al.\cite{Capaccioli} previously.
Berthier et al. have argued that although the $\xi$ from
Eq.~(\ref{xiBerthier}) is a lower bound, it should be numerically
close to the actual value of the cooperativity length. One may thus
regard this quantity as an experimentally inferred dynamical
correlation length. Fig.~\ref{xiCal} demonstrates that both the XW and
RL determined lengths are generically quite close to the experimental
$\xi$. Deviations, when present, echo those for the computed barriers
in Fig.~\ref{Cal}. This is expected based on Eqs.~(\ref{FXW}) and
(\ref{xi}). Finally, the insets in Fig.~\ref{xiCal} show the same
lengths as the main graphs, but normalized by $a$ and as functions of
temperature normalized by the respective $T_g$, to simplify
comparison.

To see whether an error in the absolute barrier values above is
potentially due to incorrectly determined bead size, we perform a
procedure analogous to what RL did to assess the effects of bead size
on their $\sigma_0$ values. This procedure gave a bead size that was
chemically reasonable (as in the As$_2$Se$_3$ example), except for
OTP\cite{RL_sigma0}, where the self-consistent bead count is too
high. (This issue was pointed out previously by Rabochiy and
Lubchenko\cite{RL_sigma0} and may be related to uncertainty in the
experimental structure factor.) Likewise, we can rescale the bead size
by a constant factor so that the hereby computed barrier values match
experiment at the glass transition temperature itself. This is
particularly straightforward in the XW case, where the barrier is
proportional to the bead count $N_b$, and so the barrier is rescaled
simply by a constant factor. It is more cumbersome to do this, but not
too much so, for the more complicated RL expressions, where
(numerically) solving a non-linear equation is required. We will call
this way of fixing the bead size, i.e., based on a predetermined value
of the absolute free energy barrier at $T_g$, the ``self-consistent''
bead count. As in Fig.~\ref{Tg}, we fix $F^\ddagger(T_g)/T_g = \ln(3.6
\cdot 10^{15}) \approx 35.8$. The bead counts obtained in this way and
the corresponding bead sizes are listed in Tables~\ref{TableRL} and
\ref{TableXW}.

\begin{figure}[h]
\centering
\subfigure{\includegraphics[width = 0.36\textwidth]{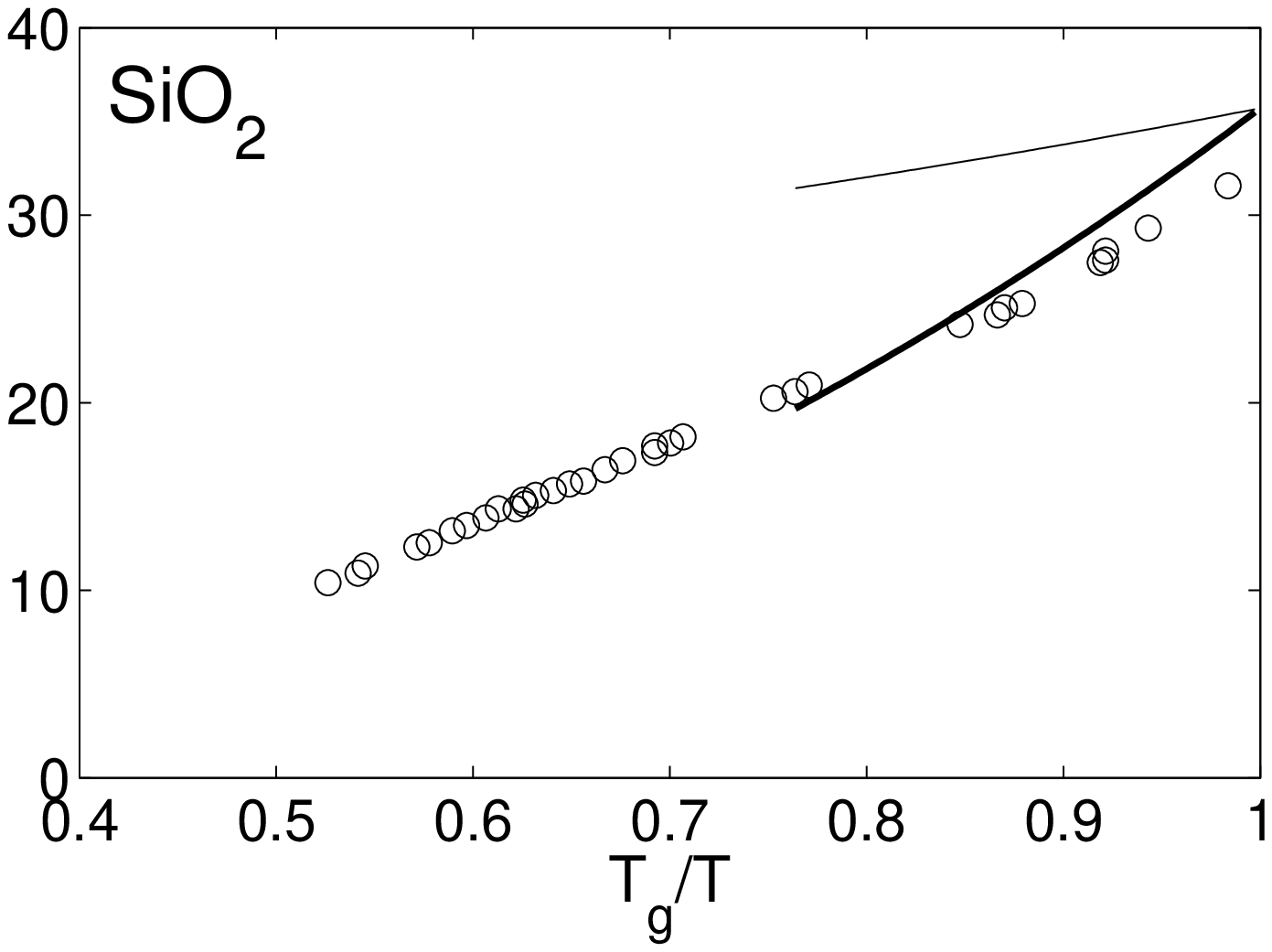}}\quad \: \: \: \: \:
\subfigure{\includegraphics[width = 0.36\textwidth]{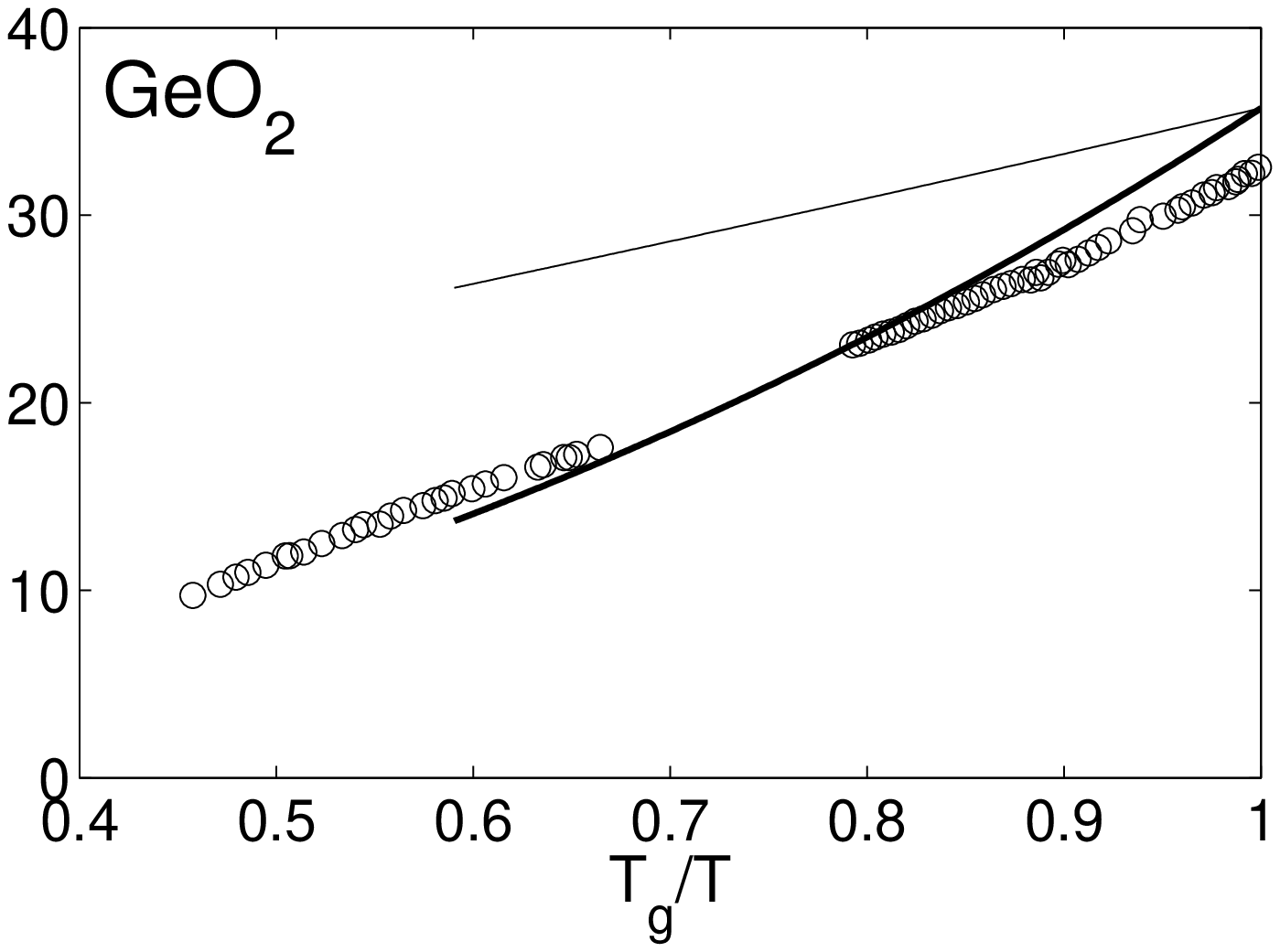}}\\
\subfigure{\includegraphics[width = 0.36\textwidth]{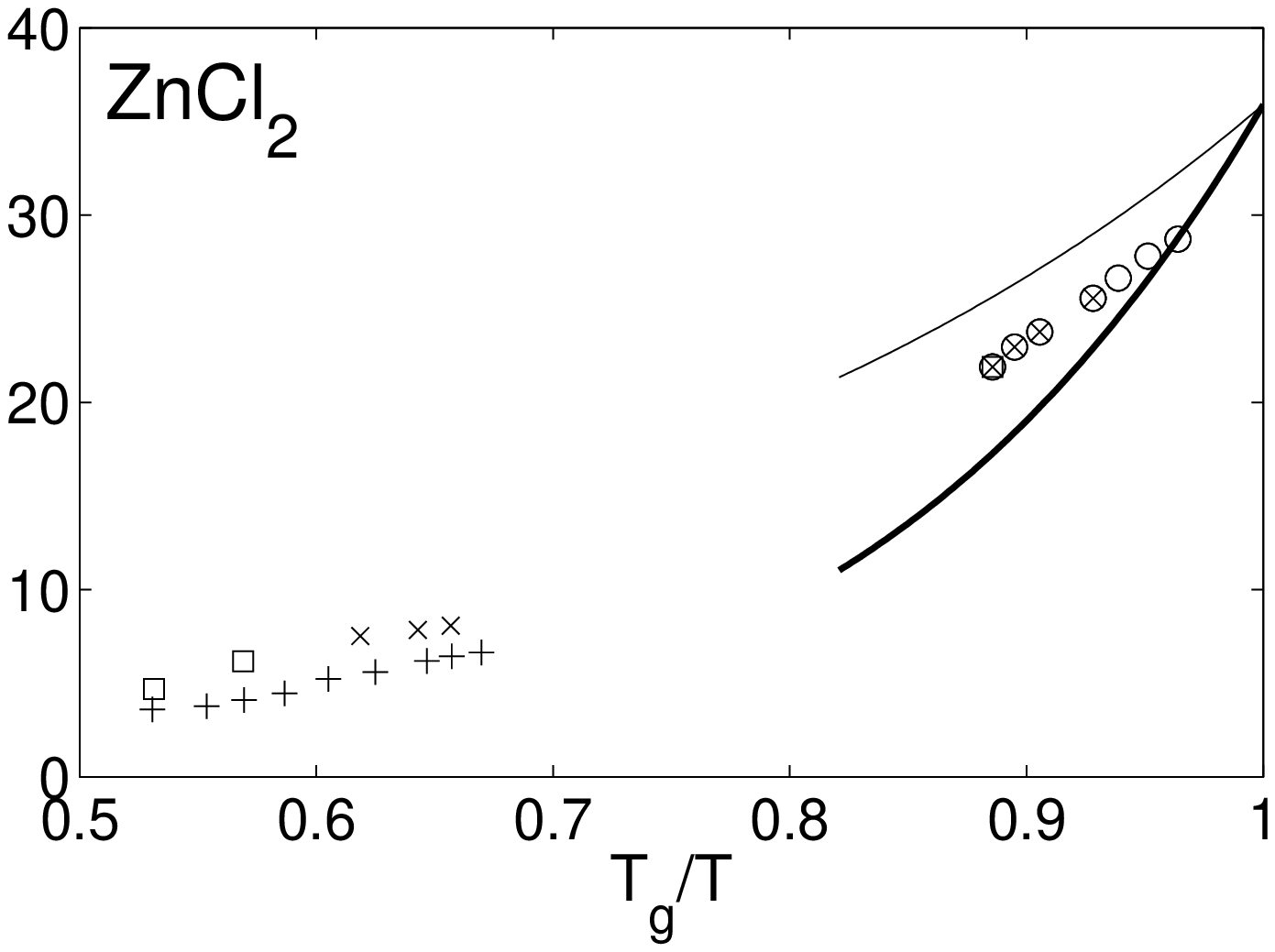}}\quad \: \: \: \: \:
\subfigure{\includegraphics[width = 0.36\textwidth]{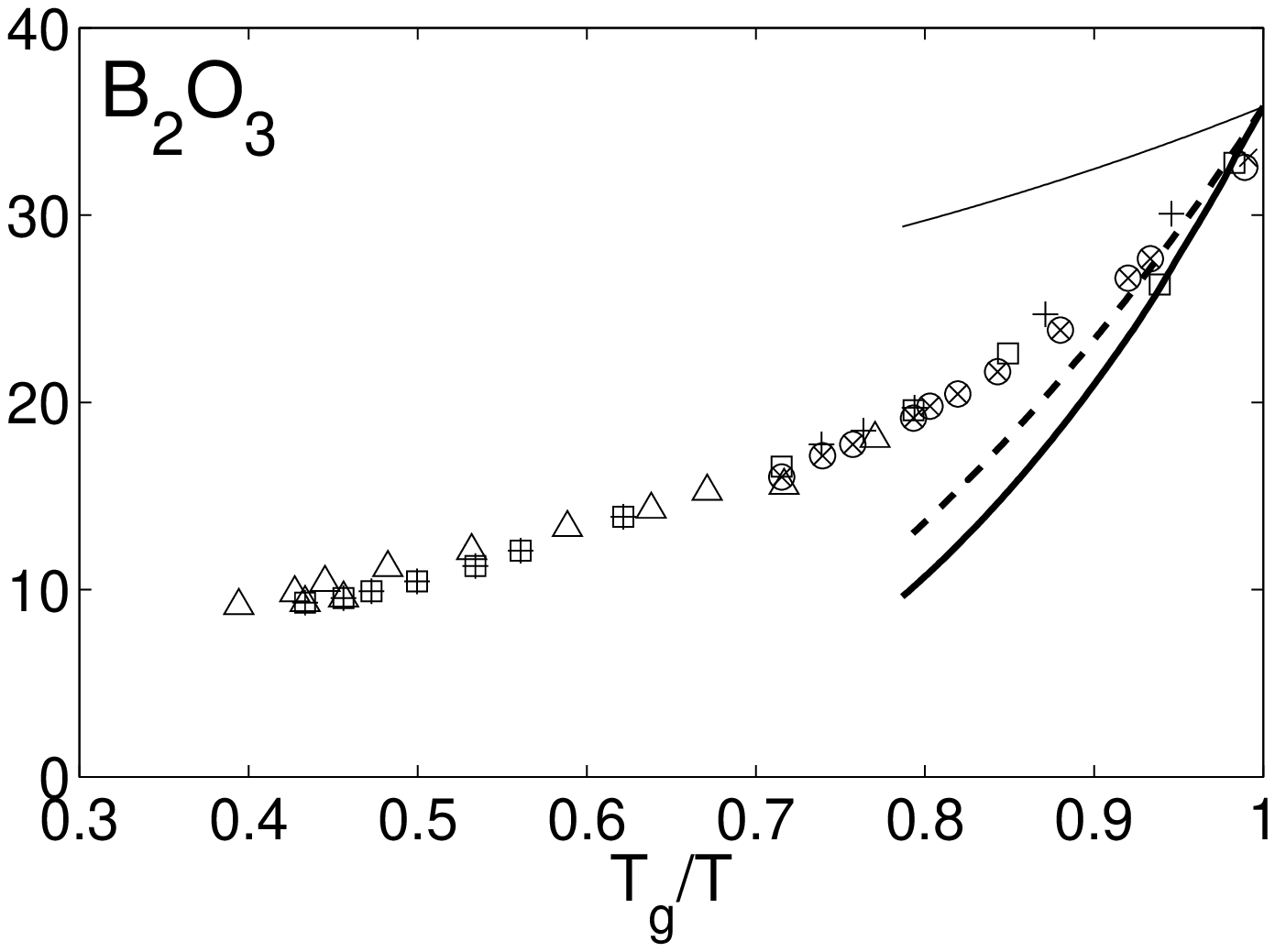}}\\
\subfigure{\includegraphics[width = 0.36\textwidth]{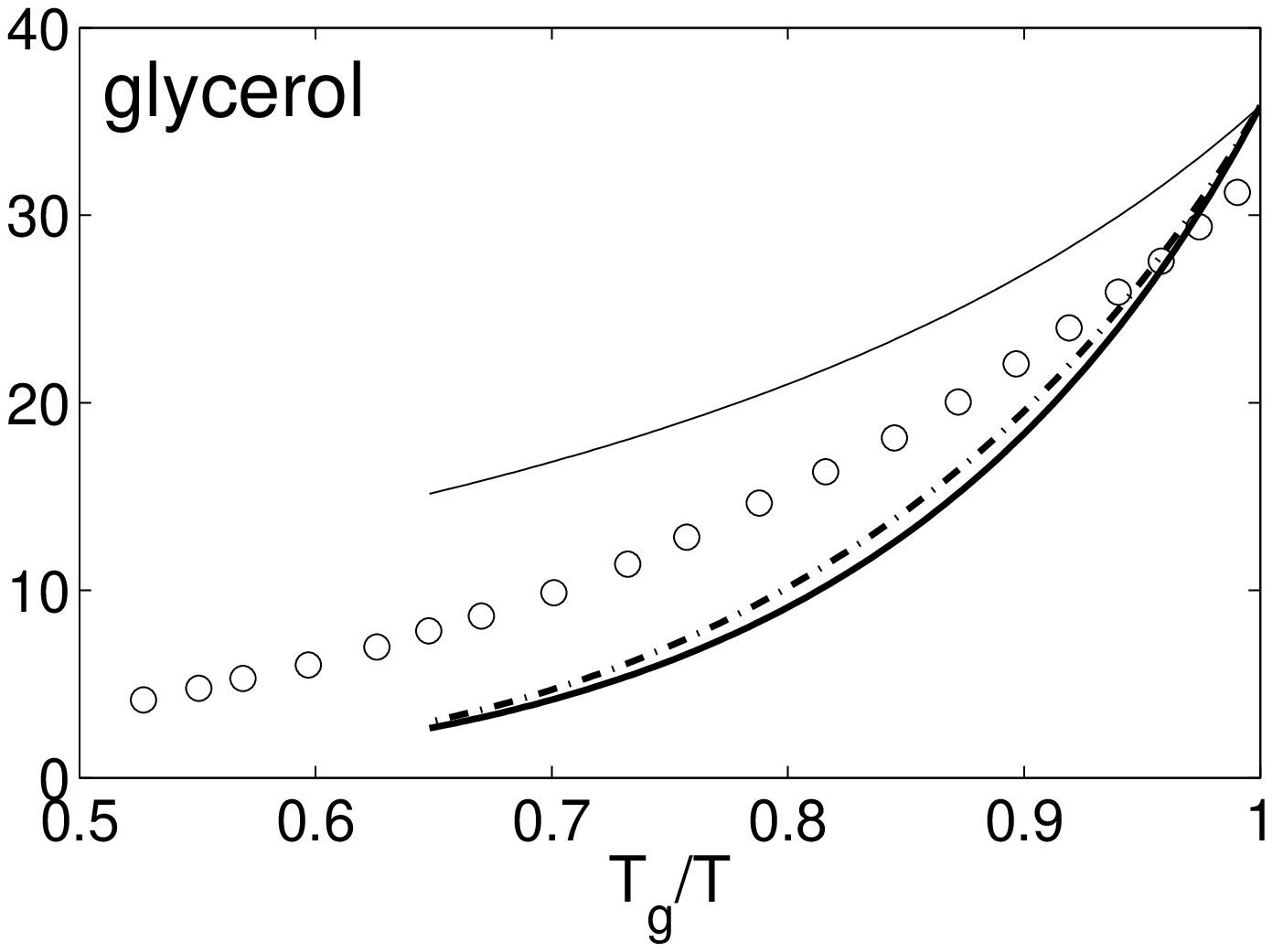}}\quad \: \: \: \: \:
\subfigure{\includegraphics[width = 0.36\textwidth]{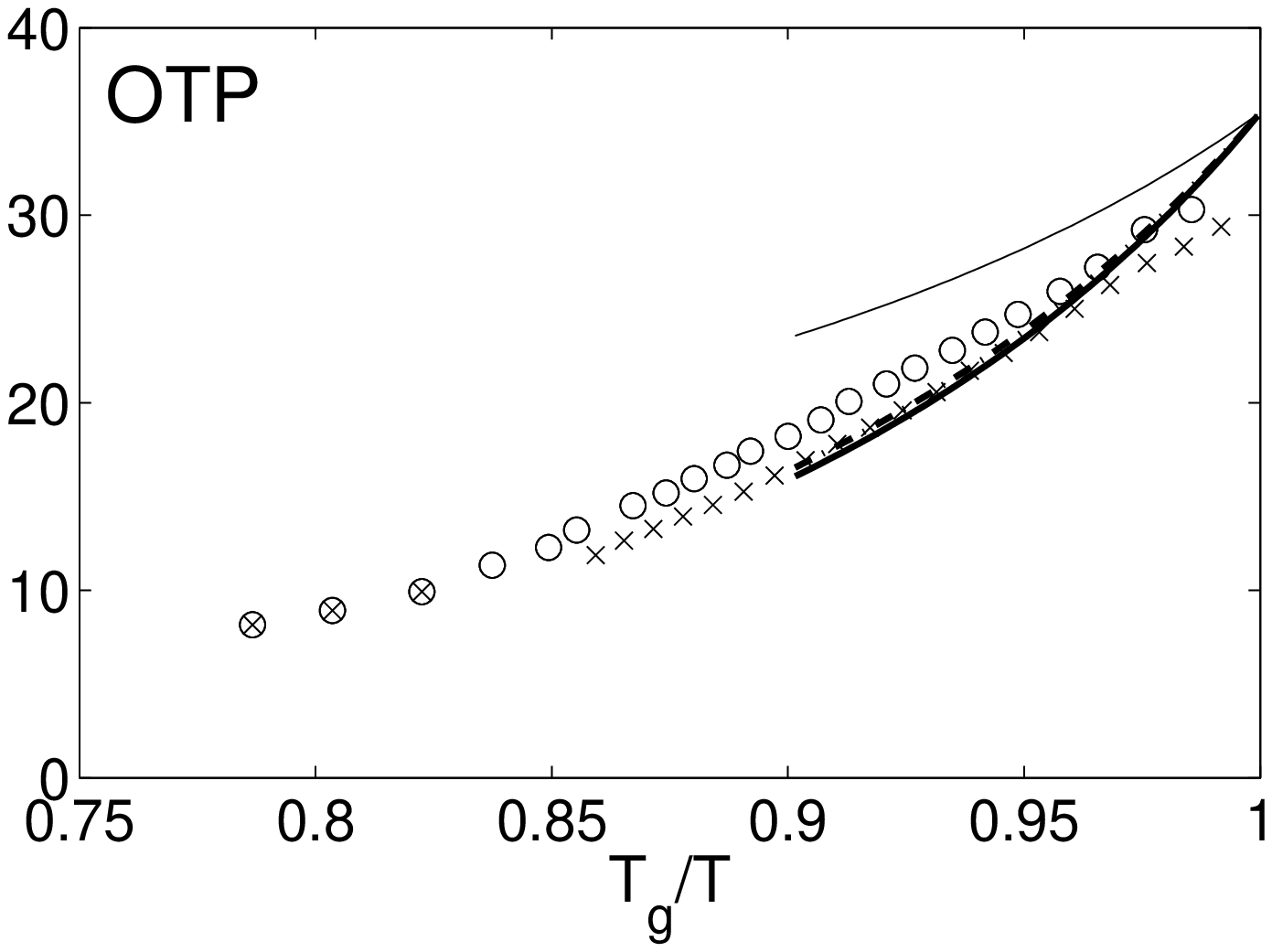}}\\
\subfigure{\includegraphics[width = 0.36\textwidth]{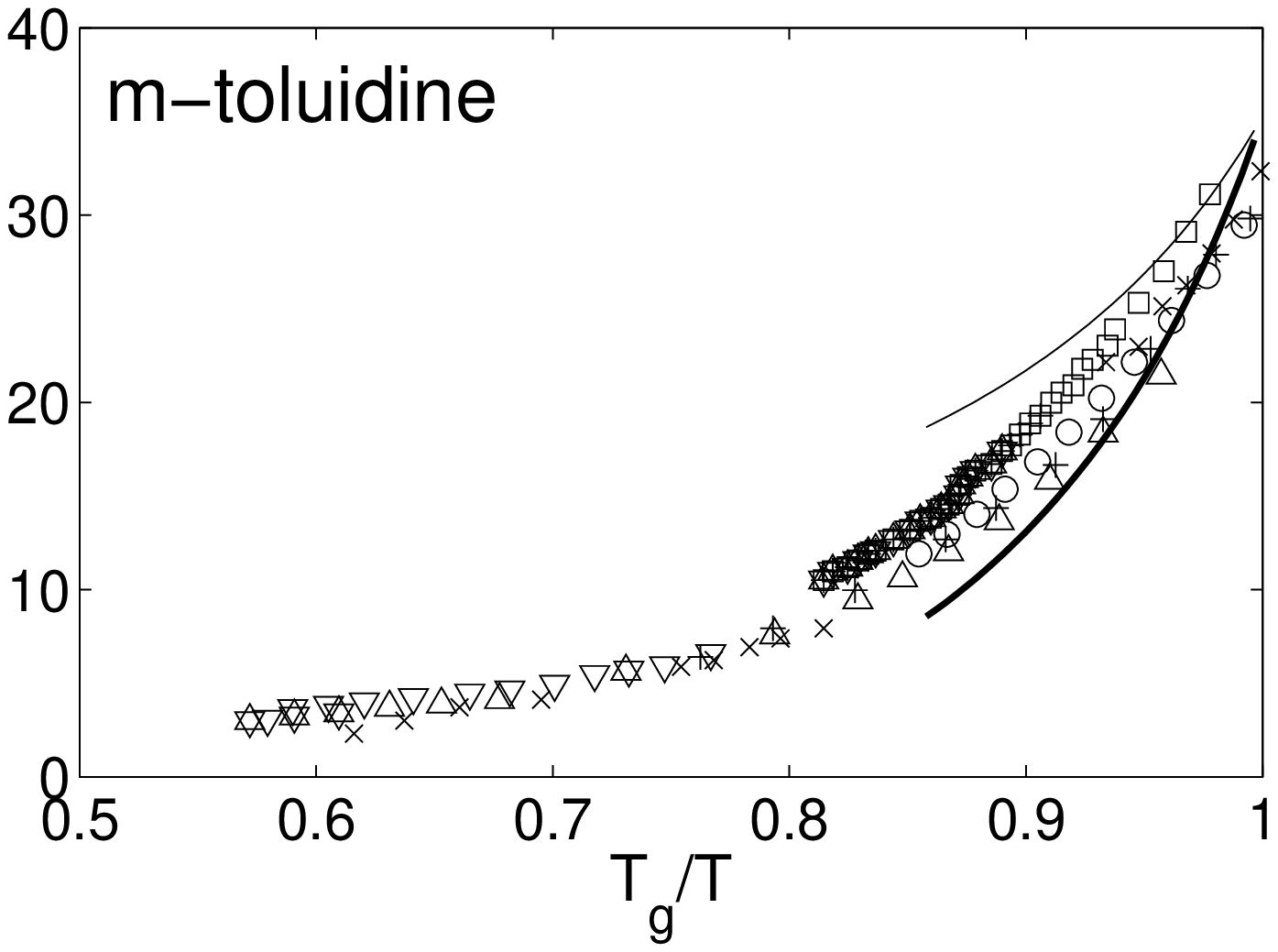}}\quad \: \: \: \: \:
\subfigure{\includegraphics[width = 0.36\textwidth]{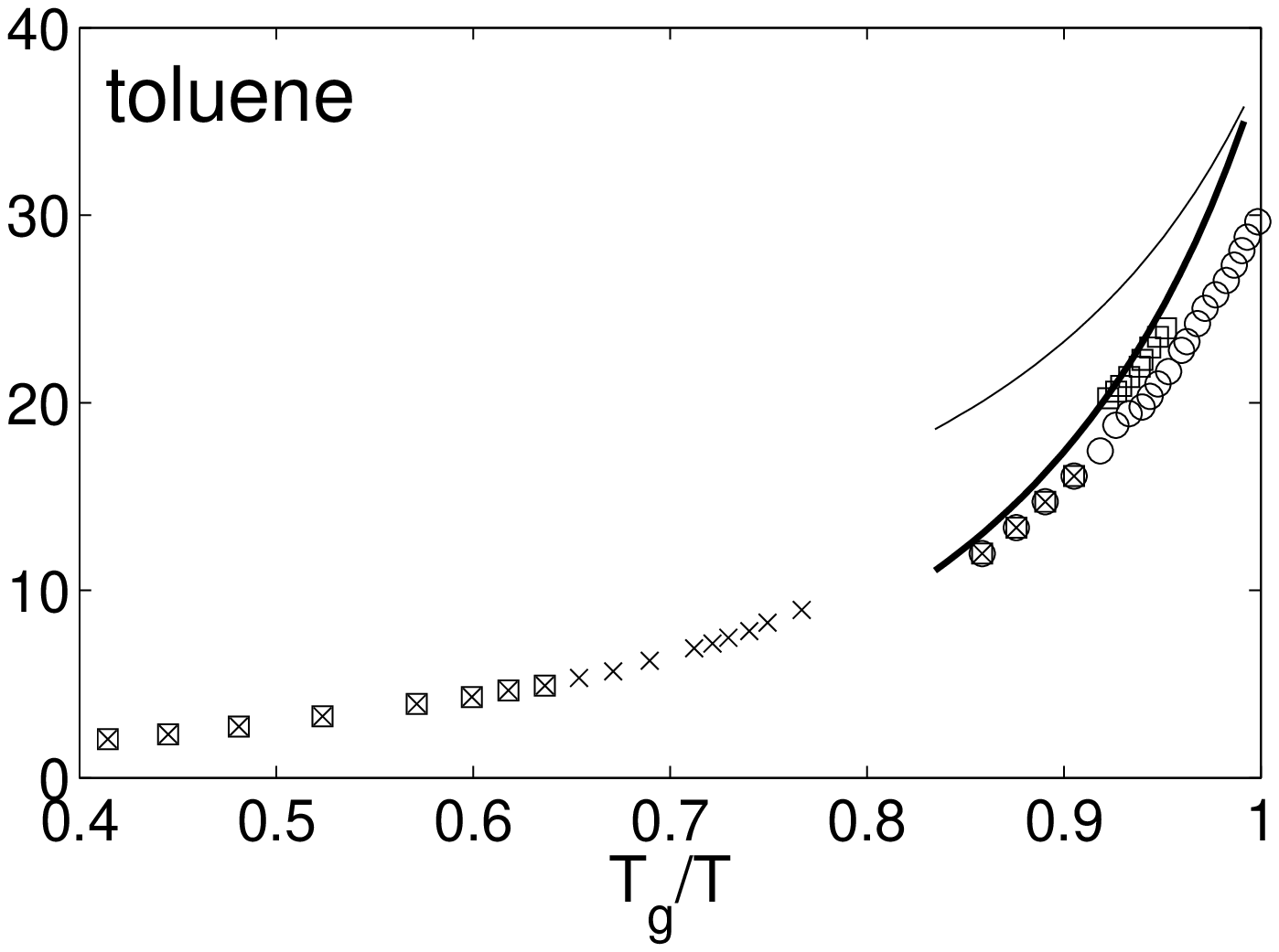}}\\
\caption{\label{SC} Barrier for $\alpha$-relaxation (divided by $k_B
  T$) as a function of temperature calculated using the
  self-consistent bead count. The XW approximation is shown with the
  thin solid line and RL approximation with thick lines, see the main
  text for detailed explanation. In all graphs, the temperature is
  scaled by the {\em experimentally} determined $T_g$. }
\end{figure}

\begin{figure}[h]
\centering
\includegraphics[width = 0.9\textwidth]{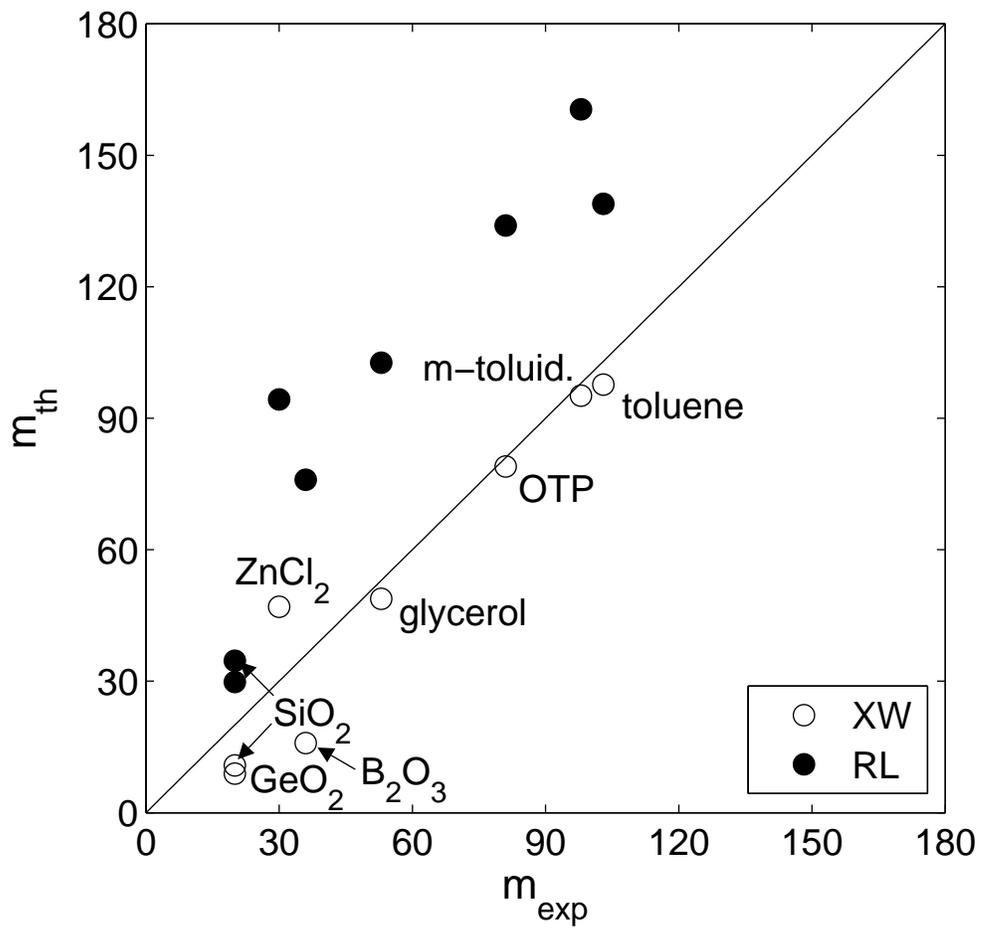}
\caption{\label{mSC} The fragility indices, Eq.~(\ref{frag}),
  corresponding to the activation barriers from Fig.~\ref{SC},
  computed using the self-consistently determined bead count, plotted
  against experimental values.}
\end{figure}

In Fig.~\ref{SC}, we present the temperature-dependent barrier values
computed when the self-consistent bead count is employed.  Neither the
high temperature relaxation time $\tau_0$ nor the time scale of the
glass transition for the experimental systems in question is
available, and so some discrepancy between theory and experiment is
built-in at the onset. This amounts to only a mild inconvenience,
however, as the reader can infer from Fig.~\ref{SC}. The latter figure
lets one focus on the {\em slope} of the temperature dependence of the
barriers, as opposed to their absolute values. The legend in this
figure is the same as in Fig.~\ref{Cal}. Now, the fragility indices
corresponding to Fig.~\ref{SC} are shown in Fig.~\ref{mSC}, to be
compared with those in Fig.~\ref{mCal}. We observe that both the
RL-based and XW-based values of the fragility index show a better
correlation with experiment than for the calorimetrically determined
bead count. For both ways to count beads, the RL-based prediction
consistently overestimates the rate of the barrier decrease with
temperature. Note that the because of the rescaling, the effect of
using a different structure factor for glycerol and OPT is quite
small.

According to Figs.~\ref{SC} and \ref{mSC}, the XW approximation
consistently yields an adequate estimate for the rate of change of the
barrier with temperature, near the glass transition as defined by a
fixed free energy barrier. At the same time, it systematically
underestimates this variation at higher temperatures. This agrees with
our earlier remarks on the progressively more dominant contribution of
the configurational entropy to the temperature variation the closer
one is to $T_K$. The RL argument suggests that some of the missing
portion of the temperature dependence in XW formulation is, in fact,
due to the temperature dependence of the elastic constants. Apart from
the overall magnitude of the barrier, the RL-predicted rates of change
of the barrier with temperature seem to be closer to experiment in the
broader temperature range, even though they are now overestimated. We
will comment on the latter in the Discussion Section. For now, it
seems instructive to assess the relative contributions of the
configurational entropy, $1/s_c(T)$, and the elastic constants,
$F^\ddagger_\text{RL} (T) s_c(T)$, to the temperature dependence of
the RL-based value of the barrier $F^\ddagger$. Note that because we
use a constant value of the structure factor $S(k)$, the latter does
not contribute to the temperature dependence of the barrier. Now, the
main panels of Fig.~\ref{dlog} show the logarithmic derivatives of the
logarithm of the entropic and elastic contributions, which allows one
to compare the corresponding rates of change of these quantities not
only for each given substance separately but also across different
substances. Here, the thick solid line depicts the entropic part. The
thin solid line corresponds to the RL prediction of the elastic
contribution alone, i.e, $F^\ddagger_\text{RL} (T) s_c(T)$. The thick
dashed line corresponds to the elastic contribution as assessed on the
experimentally determined value of the barrier, i.e,
$F^\ddagger_\text{exp} (T) s_c(T)$; this would be the actual
non-entropic contribution to the barrier change according to
Eq.~(\ref{Fr}).  We observe that the ``experimentally inferred''
elastic contribution to the temperature dependence of the barrier is
consistently lower than the contribution from the variation of the
entropy. In some cases, the so determined elastic variation is in fact
{\em negative}, which may well be due to uncertainties in the
experimental inputs. The situation with the RL-based microscopic
prediction of the elastic contribution is a bit more complicated. In
two cases, it exceeds the entropic contribution, while in the other
cases it is comparable or less than the latter. In all cases, however,
the configurational entropy becomes more important as $T_g$ is
approached. As to the inconsistency of the RL estimate, compared with
the experiment-based estimate, we suggest that our ignoring the
temperature dependence of the structure factor may be to blame. At any
rate, note that since the RL barrier overestimates the fragility, the
effect of the RL-based temperature dependence of the elastic
constants, as shown in Fig.~\ref{dlog}, is probably overestimated,
too.

\begin{figure}[h]
\centering
\includegraphics[width = 0.9\textwidth]{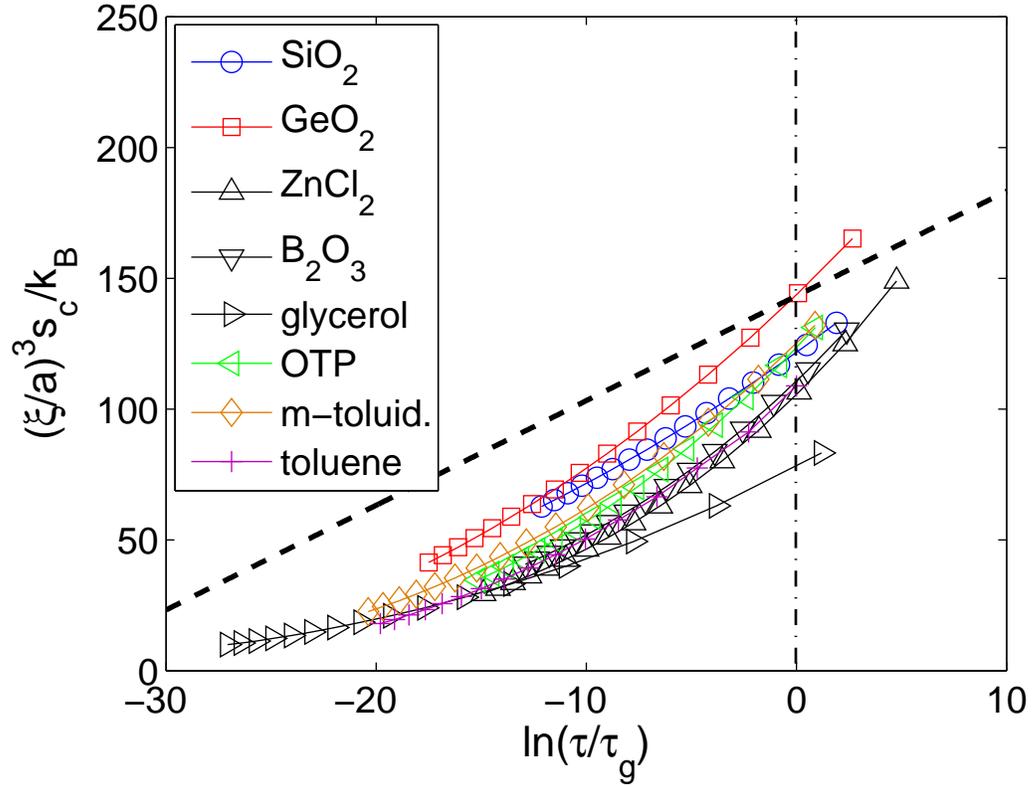}
\caption{\label{compZamp} The experimentally inferred complexity of a
  rearranging region $s_c (\xi/a)^3$ is plotted (symbols) as a
  function of $\ln (\tau/\tau_g)$, where $\tau_g$ is the relaxation
  time $\tau$ at $T_g$.  The cooperativity length $\xi$ in this graph
  is estimated as in Ref.\cite{Capaccioli}. The dashed lines line
  corresponds to the universal prediction from either form of the RFOT
  theory. This result does not depend on substance.}
\end{figure}

The above assessment of the relative contributions of the
configurational entropy and the elastic constants to the temperature
dependence of the barrier relies, of course, on the RFOT-based
expression (\ref{FXW}). One can actually test the internal consistency
of the latter RFOT expression in a manner that is independent of the
value of the surface tension $\sigma_0$, and thus is in no way
affected by the temperature dependence of the elastic constants, among
other things. There is a simple relationship between the
$\alpha$-relaxation barrier (or, equivalently, $\ln(\tau/\tau_0)$) and
the product $(\xi/a)^3 s_c /k_B$, which is often called the
``complexity'' of the rearranging region. Indeed, dividing
Eq.~(\ref{FXW}) by (\ref{xi}), one obtains that the complexity of a
rearranging region is simply the barrier, in units of $k_B T$, times a
factor of four:
\begin{equation} \label{compl} (\xi/a)^3 s_c /k_B = 4 F^\ddagger/k_B T
  = 4 \ln(\tau/\tau_0)
\end{equation}
This relationship can be interpreted as saying rearrangement involves
searching through all the states of a fixed fraction of the
rearranging region\cite{WolynesNIST}. The relationship in
Eq.~(\ref{compl}) was tested by Capaccioli, Ruocco, and
Zamponi\cite{Capaccioli} for a large number of actual substances. For
the present purposes, it is convenient to plot the complexity as a
function of $\ln(\tau/\tau_g)$, where $\tau_g \equiv \tau(T_g)$.
In Fig.~\ref{compZamp} we thus plot the complexity as a function of
$\ln(\tau/\tau_g)$ for the (experimentally determined) cooperativity
lengths $\xi$ computed according to Eq.~(\ref{xiBerthier}) and shown
earlier in Fig.~\ref{xiCal} as the dashed lines. The agreement of the
``experimentally'' determined complexity with the RFOT-based
prediction from Eq.~(\ref{compl}) is good. For one thing, the
magnitudes are nearly universal and quite consistent. The functional
dependence in Fig.~\ref{compZamp} is approximately the straight line
predicted by Eq.~(\ref{compl}), however consistently there appears to
be upward curvature. This is not entirely unexpected in view of the
aforementioned softening effects, which are not accounted for in
Eq.~(\ref{compl}), as well as mode-coupling effects\cite{BBW2008}.

\begin{figure}[h]
\centering
\includegraphics[width = 0.9\textwidth]{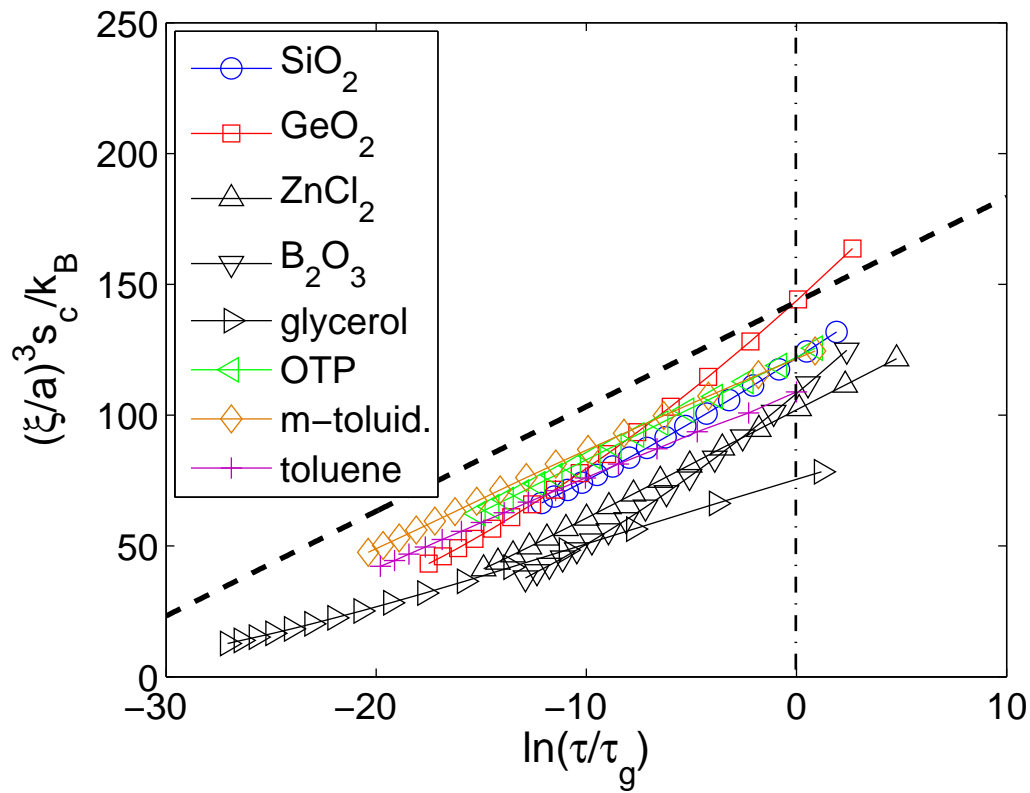}
\caption{\label{compXWLn} The complexity of a rearranging region $s_c
  (\xi/a)^3$ is plotted as a function of $\ln (\tau/\tau_g)$, where
  $\tau_g$ is the relaxation time $\tau$ at $T_g$.  The cooperativity
  length $\xi$ in this graph is estimated as in Ref.\cite{Capaccioli},
  except here we use the Xia-Wolynes-Lubchenko expression for the
  temperature dependence of the stretching exponent $\beta$ normalized
  so that it matches its experimental value at $T_g$.}
\end{figure}

Still, it seems instructive to assess the potential ambiguity stemming
from our neglect of the temperature dependence of the stretching exponent
$\beta$ when computing $\xi$ for five of the substances. Such
ambiguity is expected to be especially significant for fragile
substances\cite{XWbeta}. In the absence of experimental data, we can
utilize the expression for the temperature dependence of $\beta$
derived by Xia and Wolynes\cite{XWbeta}, based on their approximation
for the barrier distribution and later modified slightly by
Lubchenko\cite{Lionic} to achieve a better consistency between the
distribution and experimental dielectric spectra:
\begin{equation} \label{betaD} \beta_\text{XWL} = \left[ 1 + \left(
      \frac{F/k_BT}{8 \sqrt{D}} \right)^2 \right]^{-1/2},
\end{equation}
where $D$ is the fragility as defined by the Vogel-Fulcher fitting
form: $\tau = \tau_0 e^{DT_K/(T-T_K)}$.  We plot, in
Fig.~\ref{compXWLn}, the complexity values computed using the
XWL-based estimate of the temperature dependence of $\beta$ while
normalizing the $\beta(T)$'s so that their value at $T_g$ matches the
experimental values as before. For the sake of argument, we employed
the latter procedure for all eight substances including those three
for which an experimentally determined $\beta(T)$ is available. We see
utilizing a reasonable estimate for the temperature dependence of the
non-exponentiality removes much of the curvature. Clearly better
measurements of $\beta(T)$ are needed before strict values of the
scaling exponents can be assigned.

\section{Discussion}
\label{Disc}

We have presented calculations of the absolute free energy activation
barriers for $\alpha$-relaxation in supercooled liquids over a finite
temperature range from two microscopic but approximate theories both
based on RFOT ideas. We used thermodynamic, static structural, and
elastic inputs solely and no adjustable parameters. Both calculations
are based on the RFOT theory\cite{KTW, XW, LW_aging}, by which the
molecules move as a result of mutual nucleation of distinct aperiodic
free energy minima; the corresponding activation energy profile is
given by Eq.~(\ref{Fr}). Two different estimates for the mismatch
penalty $\sigma_0$, due to Xia and Wolynes\cite{XW} and Rabochiy and
Lubchenko\cite{RL_sigma0} have been employed and compared with each
other and measured kinetic data. Among its other weaknesses, the XW
estimate disregards potential effects of any temperature dependence of
the elastic constants. On the other hand, it is rather general, even
if approximate, relatively simple, and requires only thermodynamic
input information. The RL estimate, on the other hand, explicitly
accounts for details of structure and bonding in liquids. It is
however rather intricate and in its current form uses additional
measured static quantities as input.

\begin{figure}[h]
\centering
\subfigure{\includegraphics[width = 0.36\textwidth]{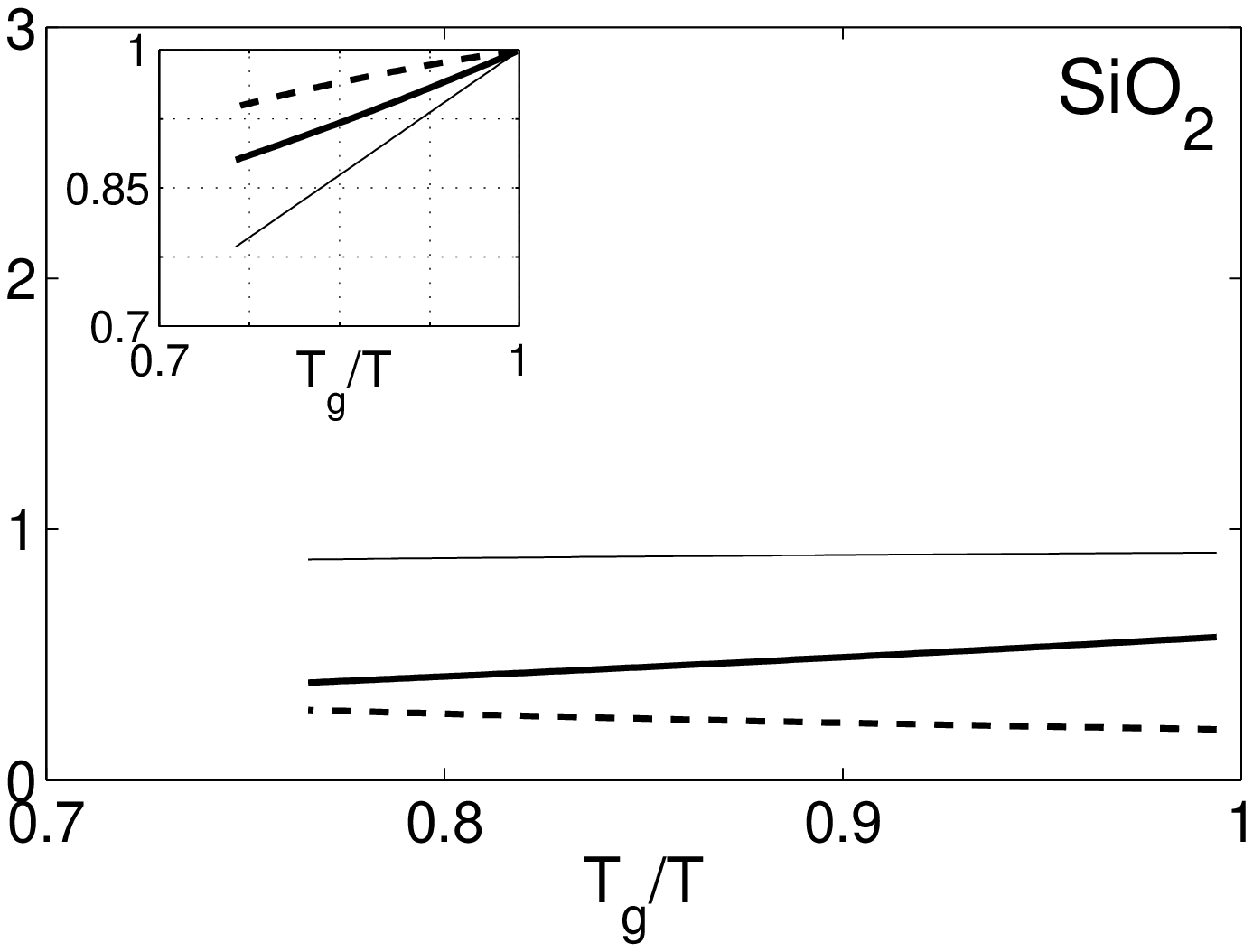}}\quad \: \: \: \: \:
\subfigure{\includegraphics[width = 0.36\textwidth]{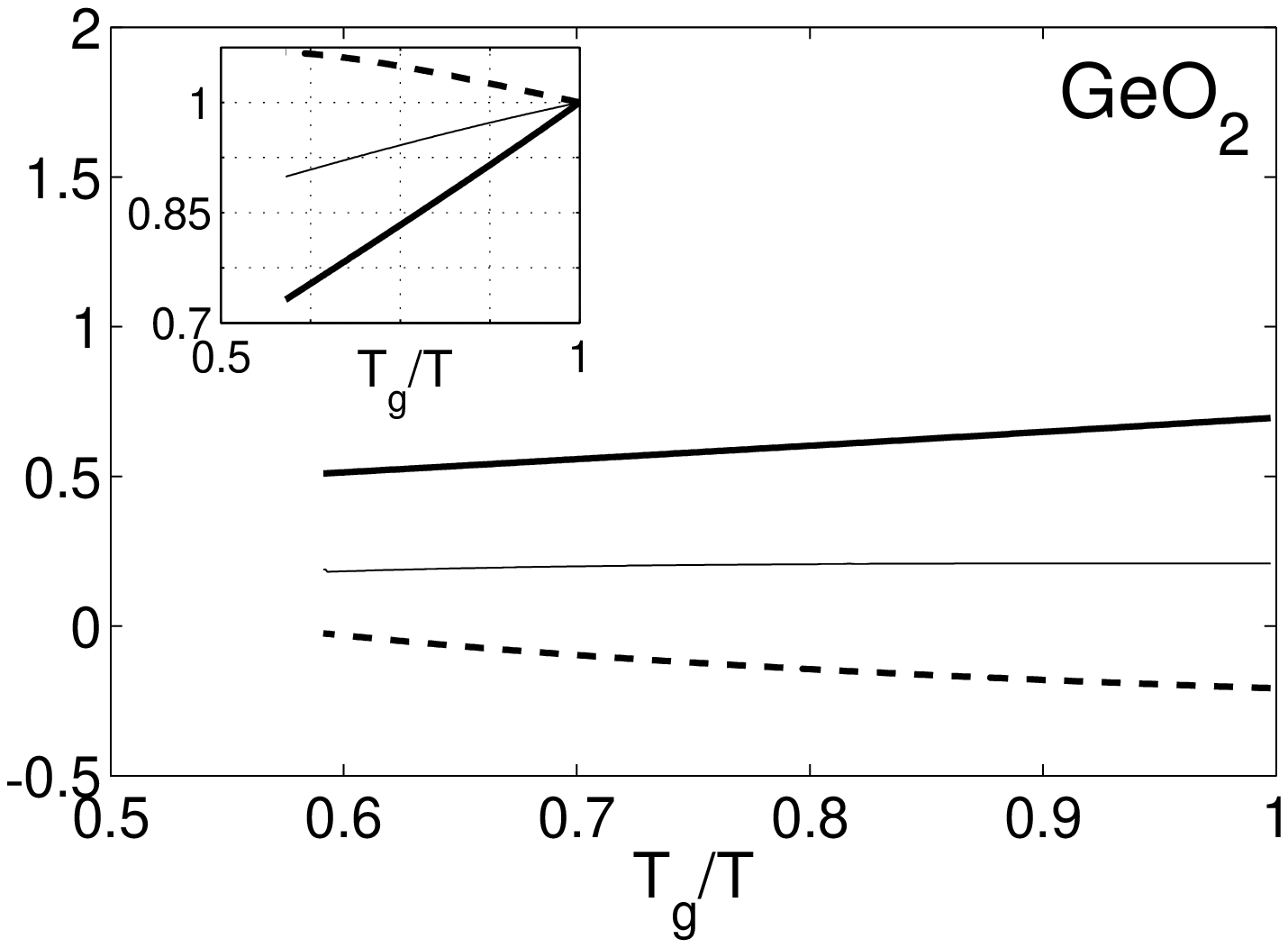}}\\
\subfigure{\includegraphics[width = 0.36\textwidth]{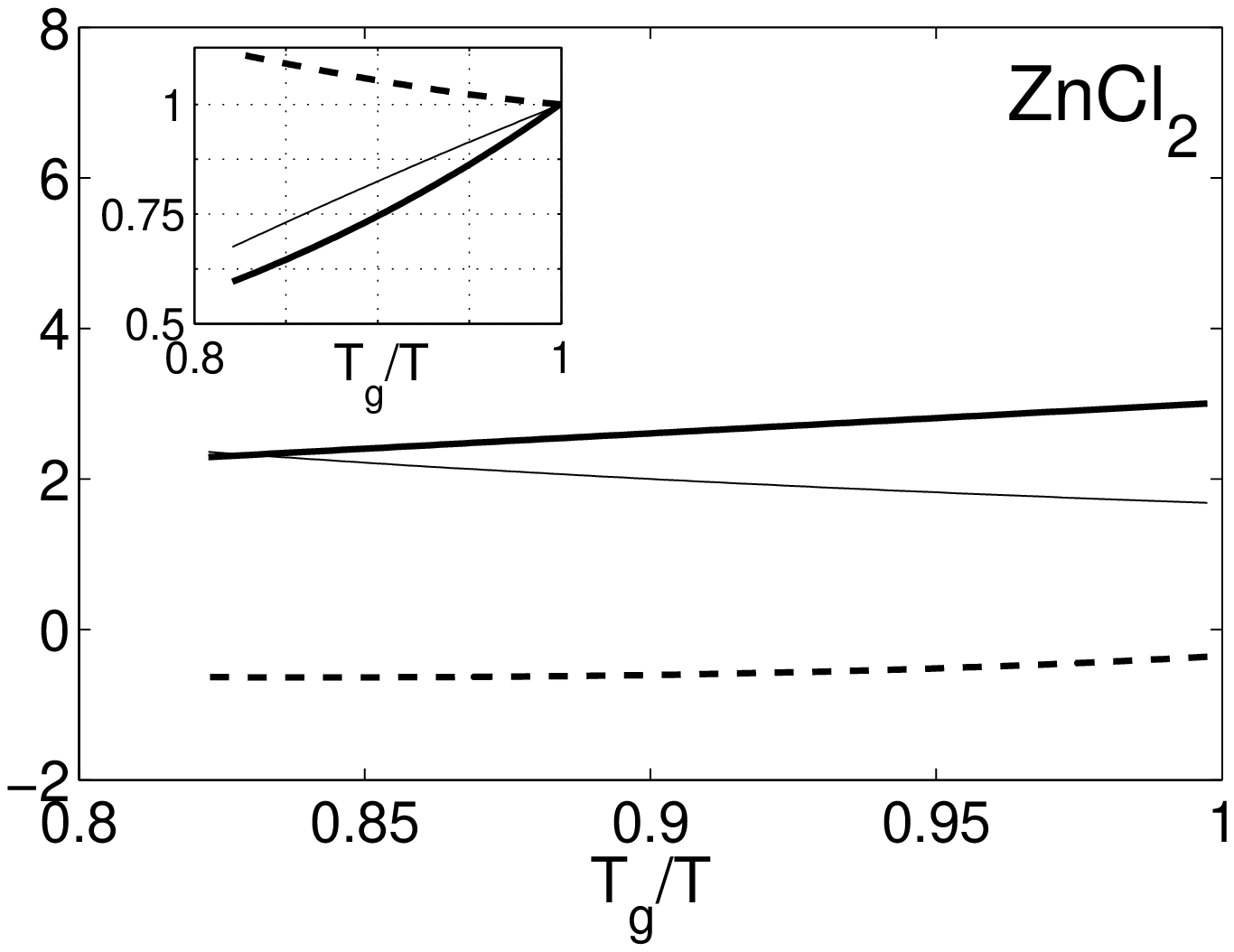}}\quad \: \: \: \: \:
\subfigure{\includegraphics[width = 0.36\textwidth]{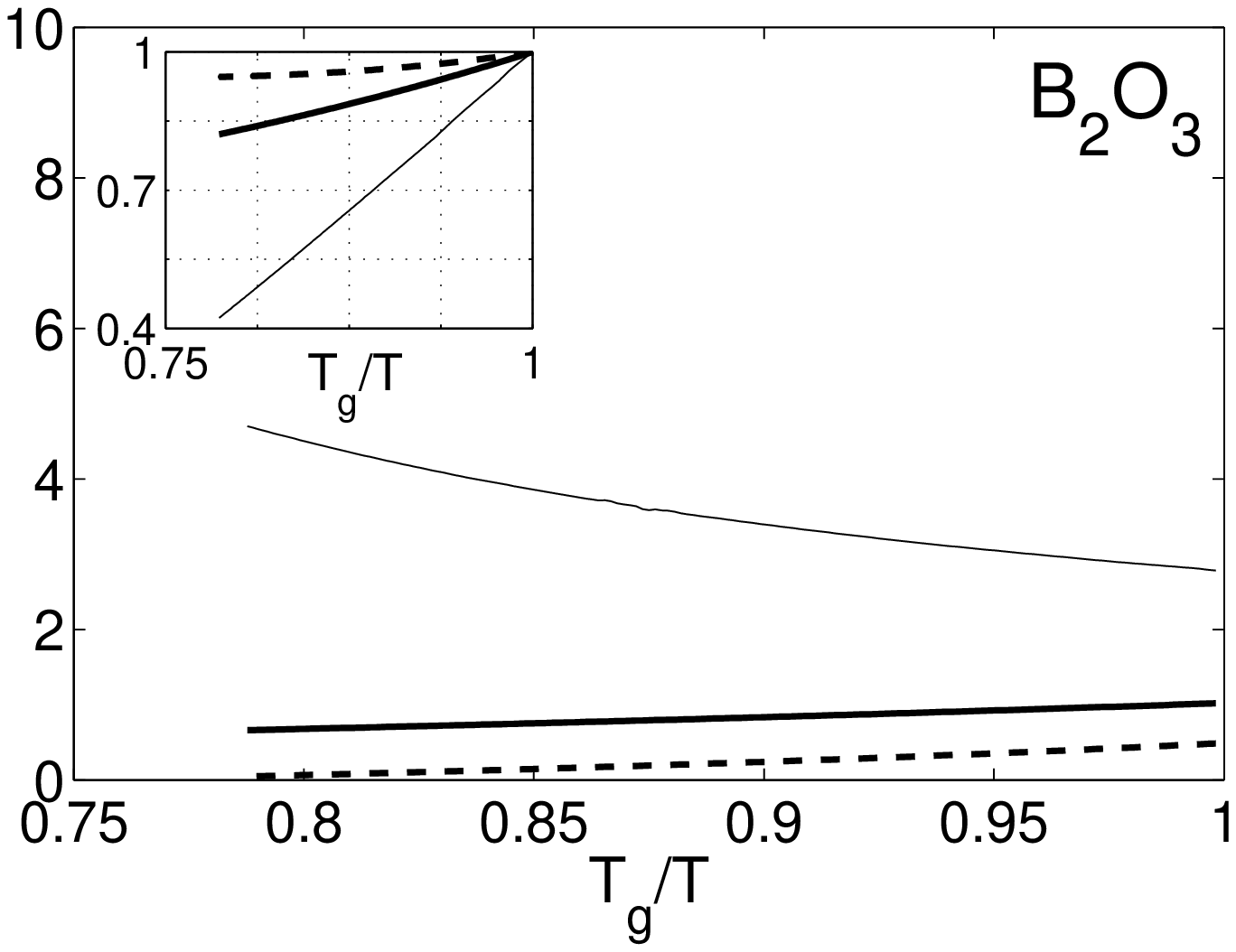}}\\
\subfigure{\includegraphics[width = 0.36\textwidth]{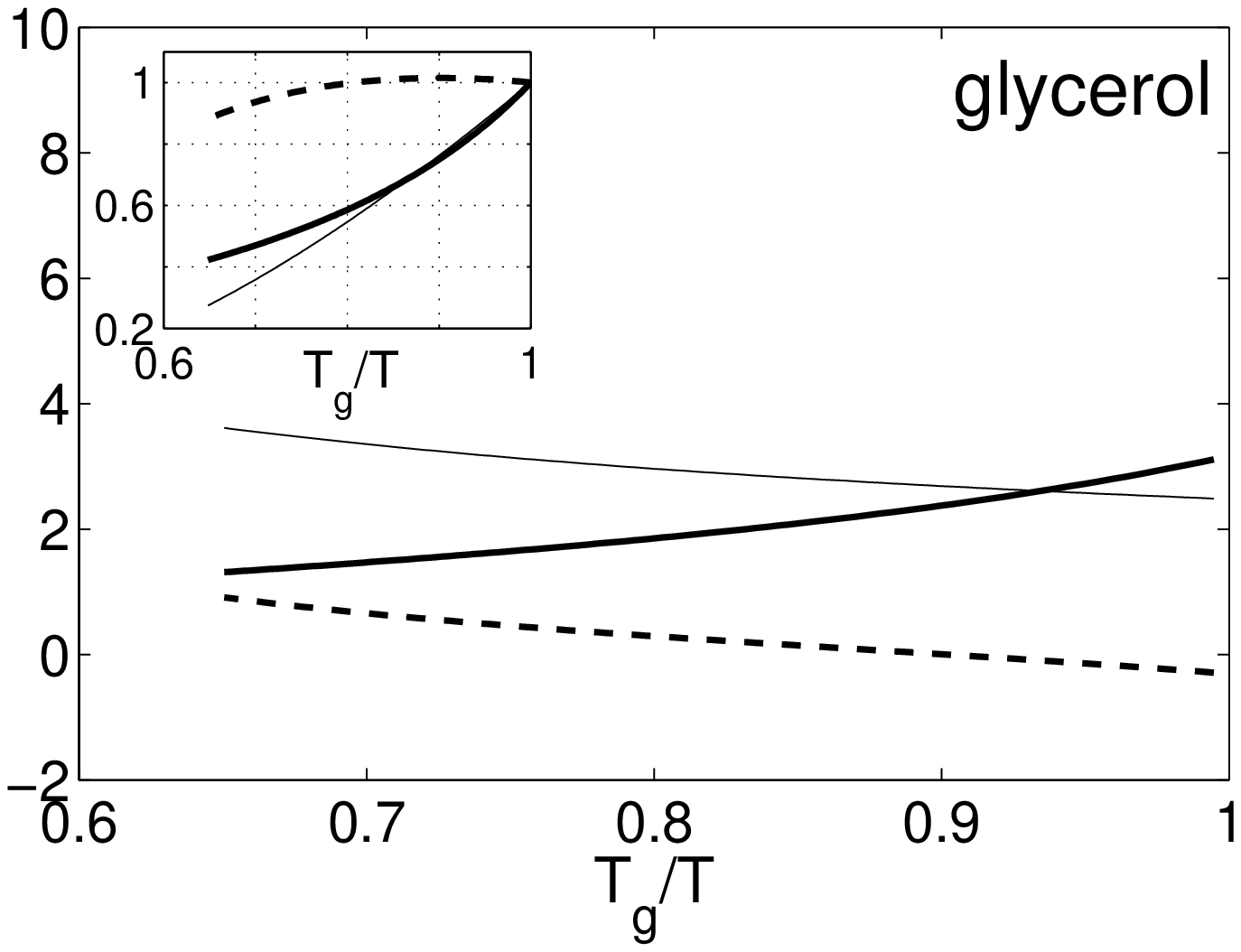}}\quad \: \: \: \: \:
\subfigure{\includegraphics[width = 0.36\textwidth]{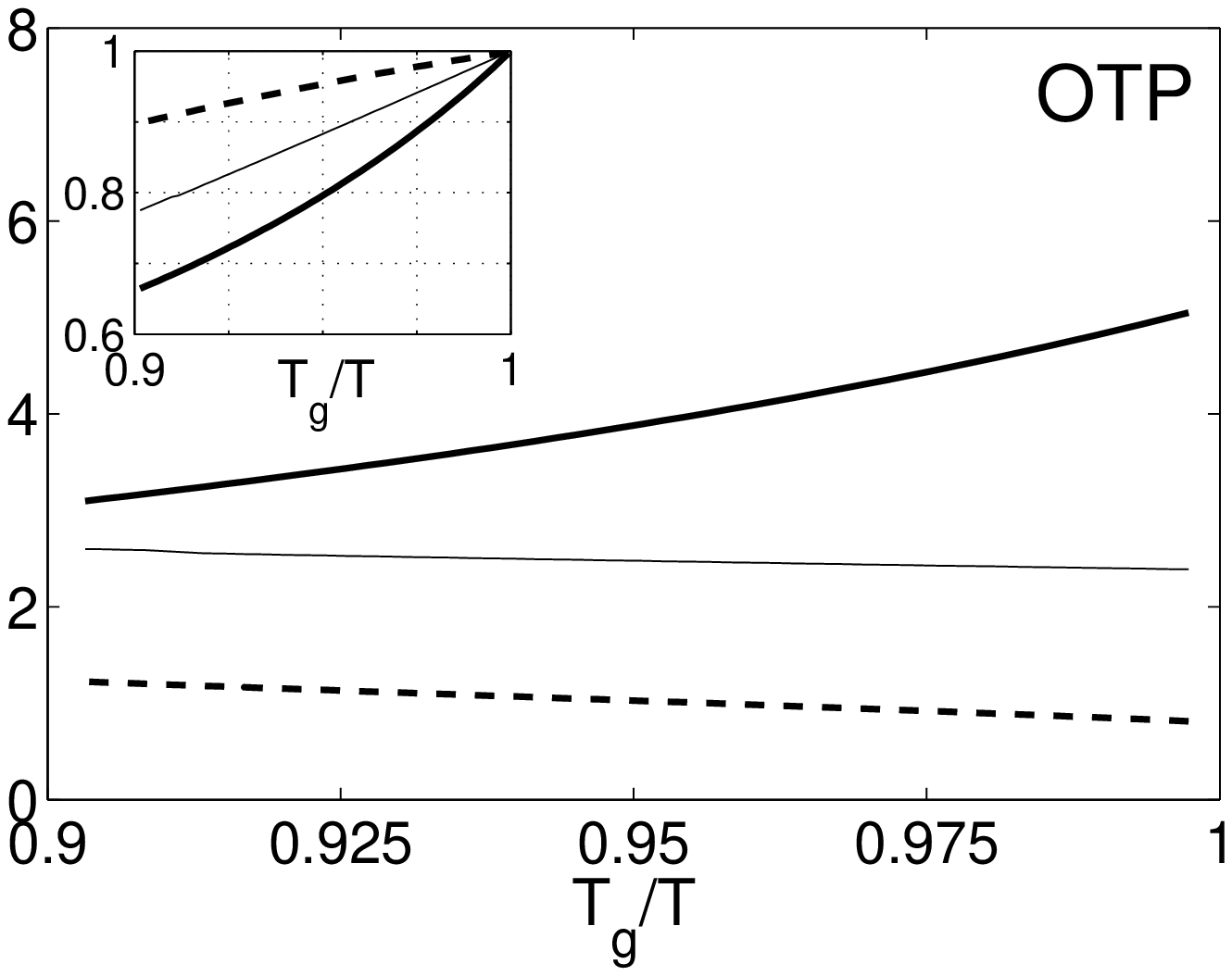}}\\
\subfigure{\includegraphics[width = 0.36\textwidth]{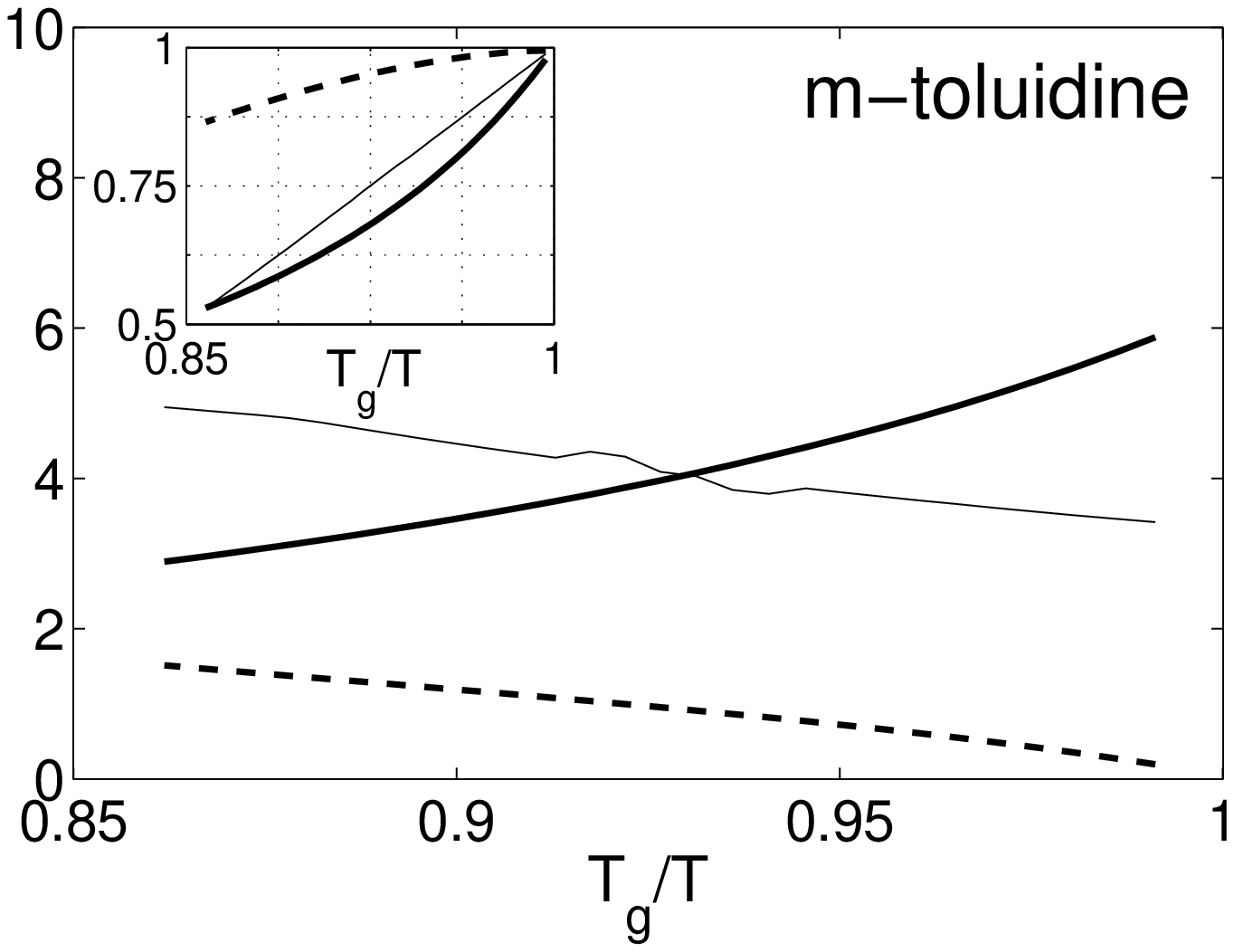}}\quad \: \: \: \: \:
\subfigure{\includegraphics[width = 0.36\textwidth]{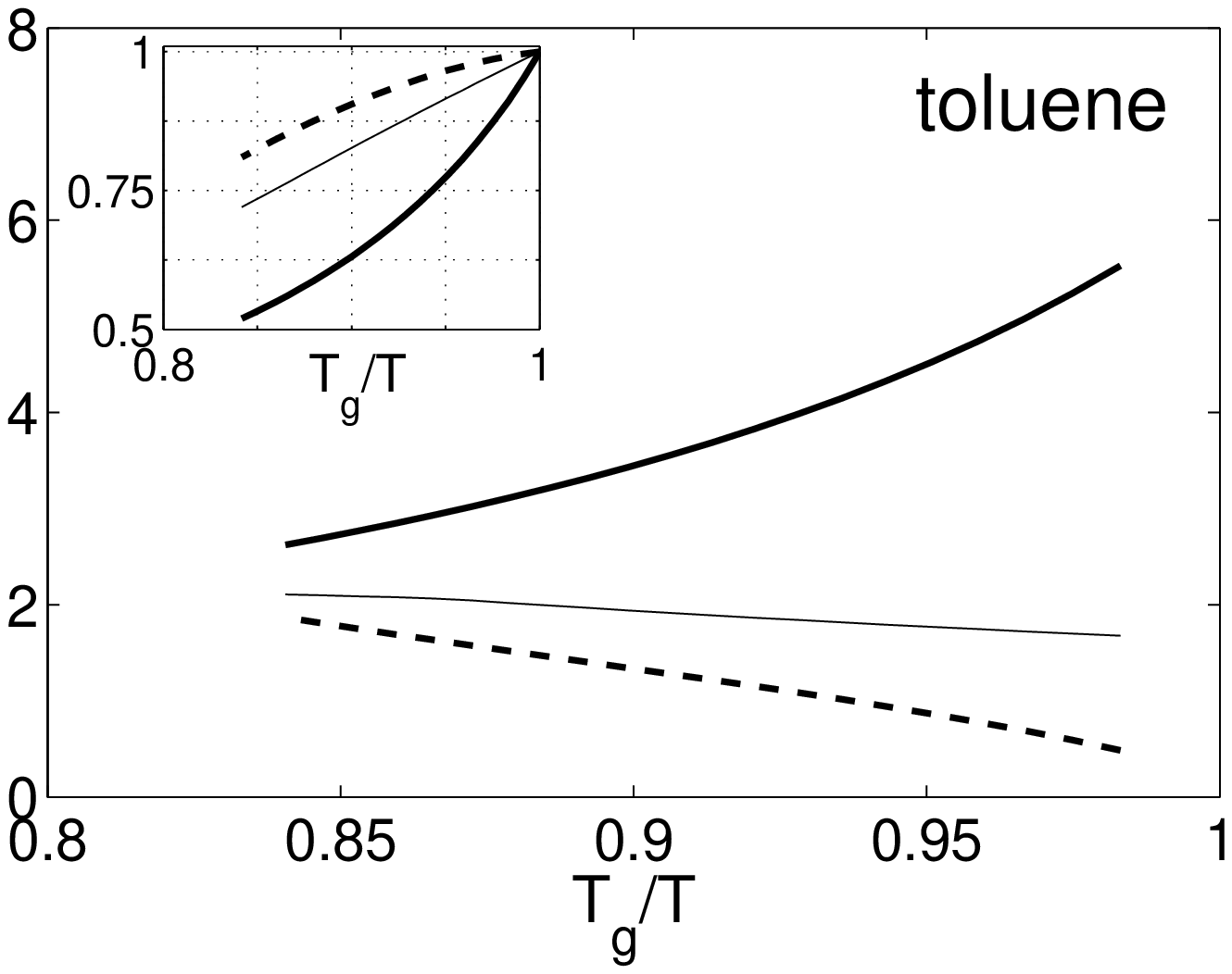}}\\
\caption{\label{dlog} The logarithmic temperature derivative,
  $(\prtl/\prtl \ln T) \ln$, of the entropic (thick solid line) and
  the remaining contribution (thin solid line) to the RL-based barrier
  $F^\ddagger$. The thick dashed line corresponds to
  $F^\ddagger_\text{exp} s_c$, where $F^\ddagger_\text{exp}$ is the
  experimentally determined barrier; this would be the actual
  non-entropic contribution to the barrier change according to
  Eq.~(\ref{Fr}). Insets: the entropic contribution ( thin line) and
  remaining contribution (thick line) to the RL-based barrier as
  functions of temperature, normalized to 1 at $T_g$. In both main
  graph and inset, we use the calorimetric bead count.
}
\end{figure}

In comparing with experiment, the XW formulation appears robustly to
determine the values of the barrier and fragility index near $T_g$,
while the RL approach seems to better capture the temperature
dependence over a more extended temperature range. The better
performance of the XW approximation near $T_g$ is perhaps to be
expected for deep reasons since the elastic effects must saturate at
very low $T$ so long as no new phase transitions occur; hereby the
dramatic barrier growth at lowering temperatures is more dominated by
the configurational entropy decrease, the lower the temperature. The
latter notion is indeed explicitly confirmed by our calculation, see
Fig.~\ref{dlog}.

The present results also indicate that first principles estimates of
the glass transition temperature itself for real substances are in
sight. Indeed, given a known value of the configuration entropy, as a
function of temperature, one can use Eq.~(\ref{Fr}), and
Eqs.~(\ref{sigmaXW}) or (\ref{sigma1}-\ref{sigma2}) to estimate the
temperature dependence of the barrier. To a specific value of the
latter, there corresponds a glass transition on the respective time
scale. Here, we have demonstrated this notion by using the
experimentally determined configurational entropy.  Note that
RFOT-based estimates of temperature scales are robust even when no
other temperature scales are explicitly used in the calculation. For
instance, Rabochiy and Lubchenko have recently estimated the crossover
temperature $T_\scr$ for several substances\cite{RL_Tcr}, which all
turned out to be consistently above the glass transition temperature.

\begin{figure}[h]
\centering
\includegraphics[width = 0.9\textwidth]{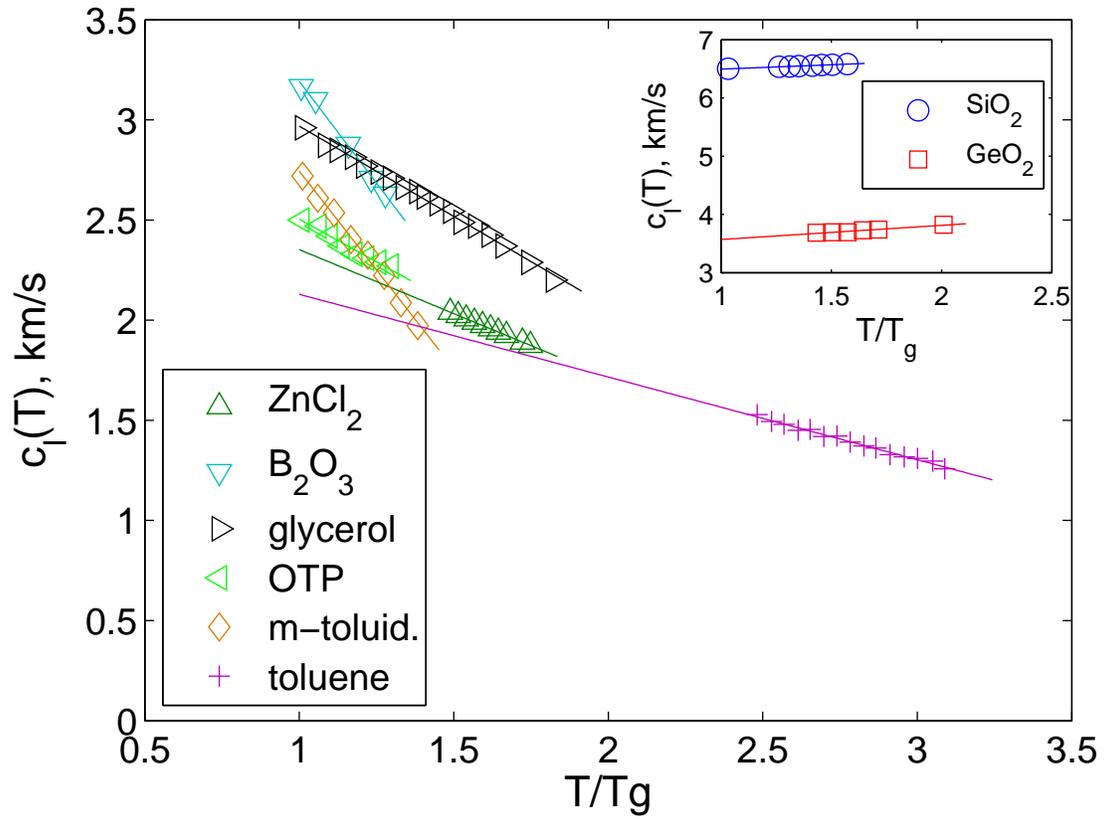}
\caption{\label{cL} Temperature dependences of the longitudinal sound
  speed. The data for two substances are shown separately in the inset
  because of their greater magnitude compared to the rest of the
  substances.}
\end{figure}

We have already mentioned that the entropy variation is the major
contributor to the barrier growth at low temperatures. Yet the
temperature dependence of factors other than the entropy, such as the
local structure, seems to be important for truly quantitative
estimates. In the RL formalism, the force details enter through the
elastic constants and the coordination, via the structure factor
$S(k)$.  Under certain circumstances, the barrier scales linearly with
the elastic constants, which hearkens back to earlier, enthalpy based
approaches to activated dynamics by Hall and
Wolynes\cite{ISI:A1987G269600055} and the so called ``shoving
model''\cite{PhysRevB.53.2171}. Note, however, that in the RL
formulation there is also another significant contribution to the
mismatch energy cost, contained in the curly brackets in
Eq.~(\ref{sigma1}). Now, the shoving model, in some of its particular
realizations\cite{dyre:224108, klieber:12A544}, proposes that the
activation barrier grows exclusively because of the increase in the
elastic constants at the ``plateau'' frequencies. In contrast, the RL
approach would seem to call for high frequency elastic constants, as
being the relevant ones for vibrational motions of individual
beads. Also importantly, the RL approach calls not for the shear
modulus alone, but the longitudinal sound speed, which amounts to a
combination $K + 4\mu/3$ of the bulk modulus $K$ and shear modulus
$\mu$. As far as these high frequency moduli are concerned, we have
seen in Fig.~\ref{dlog} that their temperature dependence is, in fact,
important for quantitative estimates of the barrier over an extended
temperature range. Still, it is essential to develop a formal theory
with regard to the latter effects, which, generally, leads to
complicated expressions as in Eq.~(\ref{sigma1}). If one were simply
to look for direct correlation between the elastic constants and the
barrier, one would discover that this correlation is complicated, if
any.  For instance, in SiO$_2$ and GeO$_2$, the measured elastic
constants {\em increase} with temperature, i.e., change in the
direction opposite to the barrier change, as we show in Fig.~\ref{cL}.

In further discussing the relative importance of the entropic
vs. elastic contributions to the barrier growth, we note that the
entropic contribution entirely, quantitatively accounts for the
discrete change in the apparent activation energy at the glass
transition\cite{LW_soft}. As a result the change is directly
correlated with the heat capacity jump, and, hence, with the
fragility\cite{XW, LW_soft, StevensonW}. A purely elasticity-based
theory would also predict a jump, if the elastic constants measured at
the plateau show a discontinuity in slope, at $T_g$, in their
temperature dependences. (High-frequency constants certainly do so.)  It thus
seems important to investigate whether the measured discontinuity of
the elastic constants would in fact correlate with the fragility of
the liquid.

The present methodology relies on the inherent relationship between
thermodynamics (the configurational entropy) and kinetics (the
activation barrier for liquid rearrangement), which has been
constructively derived within the RFOT framework, see
Eq.~(\ref{FXW}). This deep relationship is reflected in a simple,
RFOT-based prediction of a relation between the complexity of a
rearranging region and the relaxation time, Eq.~(\ref{compl}). This
relation is universal in that it does not depend on the value of the
mismatch penalty between distinct aperiodic free energy minima, which
may indeed be temperature dependent; it is therefore not subject to
various uncertainties stemming from input experimental data or
approximations needed to estimate the mismatch penalty. The complexity
also does not depend on the notion of spherical subunits, i.e.,
``beads''. We have found that the RFOT-derived, universal expression
(\ref{compl}) is indeed consistent with observation.  It is worth
remarking that although the increase of the complexity of a
rearranging region with relaxation time scale is a robust prediction
of both incarnations of the RFOT theory, it is decidedly not expected
in other pictures.  Adam and Gibbs argued for an increase in the size
of the correlated region on cooling; their argument was, however,
precisely based on the assumption of a fixed limiting complexity of a
rearranging region. This is clearly contradicted by the data when the
Berthier analysis is used: The complexity changes at least by a factor
of three over the temperature interval in question. In
enthalpically-based versions of pure elastic models the rearrangement
is assumed to have a fixed size. This assumption, combined with the
experimentally established decrease of specific configurational
entropy with increasing relaxation time, would yield a decreasing
complexity of the rearranging region upon cooling. This is
contradicted even more forcefully by the observations. The near
universality of the complexity with increasing relaxation time growth,
seen in the data, is, in our view, direct evidence that entropy and
correlation length are intimately related. This lends further, strong
support for a key non-trivial aspect of the RFOT viewpoint.

{\em Acknowledgments:} We are indebted to Matthieu Wyart for
stimulating discussions of the temperature dependence of the elastic
constants, which partially motivated this work. We thank Francesco
Zamponi and Simone Capaccioli for sharing fits of the experimental
data for the configurational entropy, relaxation times, and the
stretching exponent $\beta$; and also Philip S. Salmon for making
available $S(k)$ data for several substances.  P.R. and
V.L. gratefully acknowledge the support by the NSF Grant CHE-0956127,
the Welch Foundation Grant No. E-1765, and the Alfred P. Sloan
Research Fellowship. The work of P.G.W. is supported in part by the
Center for Theoretical Biological Physics sponsored by the NSF
(PHY-0822283) and the D.~R.~Bullard-Welch Chair at Rice University.

%

\end{document}